\documentclass[twocolumn]{autart}    

\usepackage{graphicx}          

\usepackage{setspace}
\usepackage{tikz}
\usepackage{balance}
\usepackage{graphics}
\usepackage{epsfig}
\usepackage{amsmath}
\usepackage{amssymb}
\usepackage{multirow}
\usepackage{url}
\usepackage[font=footnotesize]{subfig}
\usepackage{algorithm,color}
\usepackage[noend]{algpseudocode}
\usepackage{mathrsfs}

\usepackage{tikz}
\usetikzlibrary{arrows.meta, positioning, calc}
\usepackage{pgfplots}
\usepackage{appendix}

\newtheorem{assumption}{Assumption}
\newtheorem{remark}{Remark}
\newtheorem{theorem}{Theorem}
\newtheorem{lemma}{Lemma}
\newtheorem{definition}{Definition}
\newtheorem{scenario}{Scenario}

\DeclareMathOperator*{\diag}{diag}
\DeclareMathOperator{\Real}{\mathfrak{Re}}

\newcommand{\rank}{\mathrm{rank}}

\begin{document}

\begin{frontmatter}
\journal{arXiv.org}

\title{State Estimation Using a Network of Distributed Observers With Unknown Inputs\thanksref{footnoteinfo}} 

\thanks[footnoteinfo]{This paper was not presented at any IFAC
meeting. Corresponding author: Guitao Yang. This work has been partially supported by European Union's Horizon 2020 research and innovation program under grant agreement no. 739551 (KIOS CoE) and by the Italian Ministry for Research in the framework of the 2017 Program for Research Projects of National Interest (PRIN), Grant no. 2017YKXYXJ.}

\author[a]{Guitao Yang}\ead{guitao.yang16@imperial.ac.uk},   
\author[a]{Angelo Barboni}\ead{a.barboni16@imperial.ac.uk},
\author[a]{Hamed Rezaee}\ead{h.rezaee@imperial.ac.uk},~and 
\author[a,b,c]{Thomas Parisini}\ead{t.parisini@imperial.ac.uk}

\address[a]{Department of Electrical and Electronic Engineering, Imperial College London,
London, UK}  
  
\address[b]{Department of Engineering and Architecture, University of Trieste, Trieste, Italy}
\address[c]{KIOS Research and Innovation Center of Excellence, University of Cyprus, Nicosia, Cyprus}


\begin{abstract}                   
State estimation for a class of linear time-invariant systems with distributed output measurements (distributed sensors) and unknown inputs is addressed in this paper.
The objective is to design a network of observers such that the state vector of the entire system can be estimated, while each observer (or node) has access to only local output measurements that may not be sufficient on its own to reconstruct the entire system's state.
Existing results in the literature on distributed state estimation assume either that the system does not have inputs, or that all the system's inputs are globally known to all the observers. 
Accordingly, we address this gap by proposing a distributed observer capable of estimating the overall system's state in the presence of inputs, while each node only has limited local information on inputs and outputs.
We provide a design method that guarantees convergence of the estimation errors under some mild joint detectability conditions. This design suits undirected communication graphs that may have switching topologies and also applies to strongly connected directed graphs.
We also give existence conditions that harmonize with existing results on unknown input observers.
Finally, simulation results verify the effectiveness of the proposed estimation scheme for various scenarios.
\end{abstract}

\end{frontmatter}

\section{Introduction}
The increasing ubiquity of embedded systems has empowered sensing equipment with communication and computation capabilities that allow complex algorithms to be deployed on sensors themselves.
This is especially beneficial for larger systems comprising many different components, whose state space has a significant size or is spread over a large area.
Systems of this kind encompass smart buildings with many networked sensing points~\cite{casado2020iot} or water and power networks~\cite{bartos2018open,fadel2015survey}, where measurements are taken over a vast area.
In both cases, the centralized computation may result in additional complexity and coordination, hence running distributed algorithms is an effective design choice.
Therefore, this paper addresses the state estimation problem for a linear time-invariant (LTI) dynamical system with $N$ sensor nodes.
More generally, the \textit{distributed estimation problem} is to design a group of $N$ observers co-located with the sensors such that each observer computes an estimate of the state vector of the entire system, while only having access to measurements that are local to each node.
In general, these local measurements may not be sufficient to estimate the state, and each observer shares its own estimate with neighbouring observers over a communication network.

Many classical algorithms for state estimation, such as the Luenberger observer and the Kalman filter, have been extended in the literature in several ways for distributed state estimation.
For instance, the works \cite{olfati2007distributed} and \cite{kamgarpour2008convergence} extend the classical Kalman filter to distributed systems.
In \cite{wang2017distributed}, a general linear structure of a distributed observer is given, and no assumptions are made on the detectability of the system with respect to the individual node.
In \cite{kim2016distributed} and \cite{han2018simple}, Luenberger-like observers are designed for distributed state estimation. 
Such ideas also have been used for more complex scenarios such as resilient distributed state estimation \cite{MitraAutRob:19,MitraAut:19},
nonlinear distributed estimation \cite{YangIFAC:20,Hearxiv:19}, distributed estimation in the presence of switching topologies \cite{ugrinovskii2013distributed,XuIFAC:20}, $\mathscr{H}_\infty$-based distributed  estimation \cite{shen2010distributed}, distributed moving horizon estimation \cite{farina2010distributed}, etc.

A limitation that existing works on distributed estimation have in common is the assumption that the global system is \emph{autonomous} (i.e., there are no external inputs) or that the input information is available globally for all nodes.
However, in practice, when a system is distributed and is driven by some inputs, it may not be possible for each node to access all control signals.
Instead, each node may merely have access to its own part of the system's input, which is available locally.  
In this case, the existing distributed estimation schemes in the literature may not be effective.

In particular, the problem of distributed state estimation is still open when unknown inputs at some nodes are considered.
We aim to bridge this gap and, compared to the existing literature, the main contributions of this paper are listed below:
\begin{itemize}
    \item The nodes of the distributed observer do not have access to the entire input vector, but rather only subsets of it are assumed to be available at each node.
    
    \item The nodes exchange with their neighbours the local estimates of the whole state vector of the system, such that under certain conditions, the estimate of each node converges to the state vector of the system via a suitably designed consensus strategy.
    
    \item Under certain detectability conditions, the feasibility of the design of the proposed distributed estimation scheme is guaranteed.
    
\end{itemize}
More precisely, we propose a distributed unknown input observer (DUIO) for an LTI system with unknown disturbances, where only the information of local outputs and local inputs is accessible at each node. 
We provide rigorous (necessary and sufficient) conditions for the existence of such DUIO, depending on a rank and an appropriately defined detectability criterion.
We also show that any feasible solution of a certain linear matrix inequality (LMI) guarantees asymptotic convergence of the observers' estimates to the real state of the system.
Therefore, such LMI condition can be constructively applied to compute the gains of the DUIO, given that the existence conditions are satisfied. Furthermore, when the aforementioned detectability criterion is satisfied, the feasibility of the LMI condition is always guaranteed.

The paper is organized as follows. 
In Section \ref{sec:Prel}, some notations and basic information on graph theory are provided. The problem is formulated in Section \ref{sec:Prob}. The distributed state estimation scheme in the presence of unknown inputs at each node is proposed in Section~\ref{sec:Main1}.
Simulation results are provided in Section \ref{sec:Sim} and concluding remarks and future work are stated in Section \ref{sec:Conc}.

\section{Preliminaries}\label{sec:Prel}
Notation and some concepts and definitions of graph theory are presented in this section.
\subsection{Notation}
Throughout the paper, $\mathbb{R}$ denotes the set of real numbers and $\mathbb{C}$ denotes the set of complex numbers. 
$\mathbb{R}_{>0}$ is the set of positive real numbers. 
We partition $\mathbb C$ into $\mathbb C^- = \{s: \Real s < 0\}$ and $\mathbb C^+ = \{s: \Real s \geq 0\}$. 
$I_{n}$ stands for the ${n \times n}$ identity matrix. $\mathbf{0}_{n\times m}$ is an ${n \times m}$ all-zeros matrix. $\mathbf{1}_n$ is an ${n\times 1}$ all-ones vector. 
$|\cdot|$ stands for the standard 2-norm. 
$\otimes$ stands for the Kronecker product.
For a matrix ${A\in \mathbb{R}^{n\times m}}$, $A^\dagger$ represents the pseudo inverse of $A$ such that if $A$ is full row rank, $A^\dagger=A^\top(AA^\top)^{-1}$ and if $A$ is full column rank, $A^\dagger=(A^\top A)^{-1}A^\top$.
$\lambda_2(\cdot)$ is the second smallest eigenvalue of a real symmetric matrix.
${{\rm diag}(M_1,M_2,\ldots,M_n)}$ represents a block diagonal matrix composed of the matrices ${M_1,M_2,\ldots,M_n}$.
Similarly, $\diag_{i \in \mathcal I}(M_i)$ is a short-hand notation when the matrices are indexed by a set $\mathcal I$.
$\mathrm{Im}$ and $\mathrm{Ker}$ are respectively the image and the kernel (or null space) of a matrix. 
${\dim(\mathscr{V})}$ is the dimension of the space $\mathscr{V}$.
A `nontrivial' (sub)space $\mathscr V$ is such that $\dim(\mathscr V) > 0$. 
Moreover, if $\mathscr{R}, \mathscr{S} \subseteq \mathscr{X}$, we define the subspace $\mathscr{R} + \mathscr{S} \subseteq \mathscr{X}$ as $\mathscr{R} + \mathscr{S}= \{r+s: r\in \mathscr{R} \text{ and }  s\in \mathscr{S}\}$, and we define the subspace $\mathscr{R} \cap \mathscr{S} \subseteq \mathscr{X}$ as $\mathscr{R} \cap \mathscr{S}= \{x: x\in \mathscr{R} \text{ and }  x\in \mathscr{S}\}$. 
Accordingly, the symbol $\oplus$ indicates that the subspaces being added are independent.
We indicate that two vector spaces $\mathscr V$ and $\mathscr W$ are isomorphic by $\mathscr V \simeq \mathscr W$.
$\alpha_A(s)$ is the minimal polynomial of $A$ as follows:
\begin{equation*}
    \alpha_A(s) = \alpha_A^+(s) \alpha_A^-(s),
\end{equation*}
where the roots of $\alpha_A^+$ belong to $\mathbb C^+$, and the roots of $\alpha_A^-$ belong to $\mathbb C^-$ \cite[Chap. 3.6]{wonham1985linear}.
$\mathscr{U\!O}(C,A)$ denotes the unobservable subspace of the pair $(C,A)$ and is defined by
\begin{equation*}
\mathscr{U\!O}(C,A) = \bigcap_{k=1}^n \mathrm{Ker}\ CA^{k-1}. 
\end{equation*}
Moreover, $\mathscr{U\!D}(C,A)$ denotes the undetectable subspace of the pair $(C,A)$ and is defined by
\begin{equation*}
\mathscr{U\!D}(C,A) = \left(\bigcap_{k=1}^n \mathrm{Ker}\ CA^{k-1}\! \right) \bigcap \ \mathrm{Ker}\ \alpha_A^+(A) . 
\end{equation*}

\subsection{Graph Theory}

Communication among the observers is described by an unweighted graph ${\mathcal{G} = (\mathbf{N},\mathcal{E},\mathcal{A})}$ where $\mathbf{N}= \{1,2,\dots,N\}$  is the set of nodes (denoting  $N$ observers with local measurement), ${\mathcal{E}\subseteq \mathbf{N} \times \mathbf{N}}$ is the set of edges (denoting communication links). In the case of undirected graph, $(i,j) \in \mathcal{E}$ denotes that there exists an  edge between Node $i$ and Node  $j$, and in the case of directed graph, $(i,j) \in \mathcal{E}$ denotes an  edge from Node $i$ to Node  $j$.
Moreover, ${\mathcal{A}=[a_{ij}] \in \mathbb{R}^{N\times N}}$ denotes the adjacency matrix. If the graph is undirected, $a_{ij}=a_{ji} =1$  if ${(i,j)\in \mathcal{E}}$, and $a_{ij}=a_{ji} =0$ otherwise, while in the case of directed graph, $a_{ij}=1$  if ${(j,i)\in \mathcal{E}}$, and it is zero otherwise. In this condition, Node $j$ is a neighbor of Node $i$ if $a_{ij}=1$.
An undirected graph is  \textit{connected} if there exists a path of edges connecting each pair of its nodes. Moreover, a directed graph is  \textit{strongly connected} if there exists a path in each direction connecting each pair of its nodes. 

The Laplacian matrix associated with the graph $\mathcal{G}$ is a  matrix ${\mathcal{L}=[\ell_{ij}]\in \mathbb{R}^{N\times N}}$ described as
\begin{equation*}
\ell_{ij}=\left\{\begin{array}{ll}
\sum_{j=1,i\neq j}^{N}a_{ij}&\quad i=j\\
-a_{ij}&\quad i \neq j,
\end{array}
\right.
\end{equation*}
where has rows with zero entries summation. Therefore, $\mathcal{L}$  always has a zero eigenvalue, and if $\mathcal{G}$ is connected or strongly connected, all the other eigenvalues are on the open right half plane. If the graph is undirected, $\mathcal{L}$  is also symmetric whose all the nonzero eigenvalues are real \cite{RenCSM:07}. In this condition, $\lambda_2(\mathcal{L})$ denote the
 \textit{algebraic connectivity} of the graph \cite{OlfatiTAC:04}.
 
The graph $\mathcal{G}$ is called balanced, if for all $i\in \mathbf{N}$, $\sum_{i=1}^{N}a_{ij}=\sum_{j=1}^{N}a_{ji}$. If $\mathcal{G}$ is connected or strongly connected and balanced, the right and left eigenvectors associated with the zero eigenvalue are ${\mathbf{1}_N/\sqrt{N}}$ \cite{RenCSM:07}.  Moreover, the Laplacian matrix associated with any balanced graph is positive semidefinite~\cite{OlfatiTAC:04}.

\section{Problem Statement}\label{sec:Prob}
Consider the dynamical system described as
\begin{equation}\label{eq:system}
    \dot{x} = Ax + Bu + Dw,
\end{equation}
where ${x \in \mathbb{R}^n}$ represents the state vector, ${u\in \mathbb{R}^m}$ is the control input, $w \in \mathbb{R}^q $ is an unknown external disturbance, ${A\in \mathbb{R}^{n\times n}}$ is the state matrix, ${B\in \mathbb{R}^{n\times m}}$ denotes the input matrix, and $D \in \mathbb{R}^{n\times q}$ is the disturbance matrix gain. 
We assume that the outputs of the system are measured via a distributed measurement system comprising of a group of sensors distributed over $N$ nodes, namely   
\begin{equation}
    \label{eq:output}
    y_i = C_ix,
\end{equation}
with $C_i \in \mathbb{R}^{p_i\times n}$.

In order to further distinguish the locally available signals, we partition the system's inputs into a component $u_i$ -- that is local to and assumed to be known at Node $i$, that is assumed to be known by the observer -- and a  component $\acute{u}_i$ -- that is unknown and can instead be assimilated to an exogenous disturbance.
In symbols, we then have
\begin{equation*}
    \label{eq:inputdecomp}
    B u = B_i u_i +\acute{B}_i \acute{u}_i,
\end{equation*}
where $u_i \in \mathbb{R}^{r_i}$, $B_i \in \mathbb{R}^{n\times r_i}$, $\acute{u}_i \in \mathbb{R}^{d_i}$, and $\acute{B}_i \in \mathbb{R}^{n\times d_i}$, with $r_i+d_i=m$.
Then, as $w$ is also unknown, we define 
$$\acute{w}_i = \begin{bmatrix}\acute{u}_i^\top &w^\top\end{bmatrix}^\top$$ 
as the locally unknown inputs and $\bar{B}_i=\begin{bmatrix}\acute{B}_i &D\end{bmatrix}$ as the known gain of the unknown terms. 

\begin{assumption}\label{as:CB} 
    The matrix $\bar{B}_i$ is full column rank for all $i \in \mathbf{N}$.\hfill $\triangleleft$
\end{assumption}

\begin{remark}\em
Note that Assumption~\ref{as:CB} does not cause any loss of generality and is typically made in the literature of estimation with unknown disturbances \cite{chen1996design}.
In fact, it is always possible (by means of singular value decomposition, for instance) to decompose $\bar B_i$ in a product $\bar B_i = \bar B_i' \bar B_i''$, where $\bar B_i'$ is full column rank and $\bar w_i' = \bar B_i'' \acute w_i$ constitutes the new unknown input. \hfill $\triangleleft$
\end{remark}

As the objective is to reconstruct the state vector $x$, we consider a distributed observer $\mathcal O = \{ \mathcal O_i \}_{i \in \mathbf N}$ comprising of $N$ local nodes (or observers) $\mathcal O_i$ located at each sensor node, where each observer has access to just its local outputs $y_i$ and local inputs $u_i$.
Furthermore, the local observers are connected over a communication network that lets them exchange their state estimates.

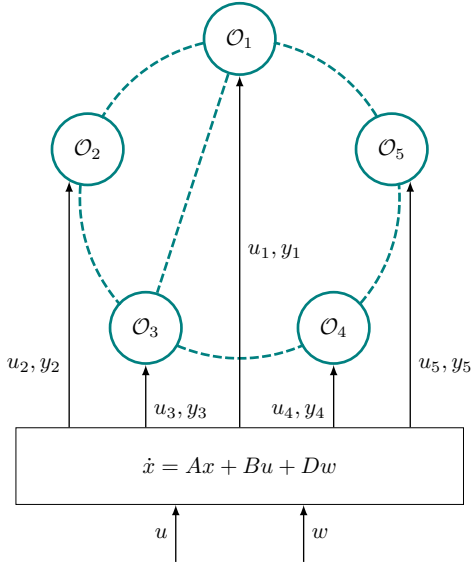
\begin{figure}
    \centering
    \scalebox{0.85}{\begin{tikzpicture}
\def\off{18}
\def\N{5}
\def\R{2.5}
\pgfmathparse{360/\N}
\edef\step{\pgfmathresult}

\colorlet{net}{teal}
\tikzset{comm/.style = {color=net, very thick, dash pattern=on 4pt off 1.5pt}}
\coordinate (O) at (0,0);
\foreach \x in {1,...,\N} {%
    \node [circle, 
        draw, 
        color=net, 
        fill=white, 
        text=black, 
        very thick,
        inner sep=6pt] (N\x) at (\x*\step+\off:\R) {$\mathcal O_{\x}$};
}
\pgfmathanglebetweenpoints{
    \pgfpointanchor{N1}{west}}{
    \pgfpointanchor{O}{center}}
\edef\anglestart{\pgfmathresult}
\pgfmathanglebetweenpoints{
    \pgfpointanchor{N2}{north}}{
    \pgfpointanchor{O}{center}}
\edef\angleend{\pgfmathresult-0.7}
\foreach \x in {1,...,\N} {
    \edef\rotation{\step*(\x-1)}
    \draw[comm,rotate=\rotation] (\anglestart-180:\R) arc[radius = \R, start angle = \anglestart-180, end angle = \angleend-180];
}
\draw [comm] (N1) -- (N3);

\node [draw,
        rectangle,
        below=2.2*\R of N1,
        minimum width=7cm,
        minimum height=1.2cm] (sys) {$\dot x = Ax + Bu + Dw$}; 

\draw [-latex, semithick] (sys.north -| N1) -- (N1) node[midway, right] {$u_1, y_1$};
\draw [-latex, semithick] (sys.north -| N2.240) -- (N2.240) node[near start, left] {$u_2, y_2$};
\draw [-latex, semithick] (sys.north -| N3) -- (N3) node[near start, right] {$u_3, y_3$};
\draw [-latex, semithick] (sys.north -| N4) -- (N4) node[near start, left] {$u_4, y_4$};
\draw [-latex, semithick] (sys.north -| N5.300) -- (N5.300) node[near start, right] {$u_5, y_5$};

\coordinate (bot) at ($(sys)+(0,-1.5)$);
\draw [-latex, semithick] ([shift={(1,0)}]bot) -- ([shift={(1,0)}]bot |- sys.south) node[midway, right] {$w$};
\draw [-latex, semithick] ([shift={(-1,0)}]bot) -- ([shift={(-1,0)}]bot |- sys.south) node[midway, left] {$u$};
\end{tikzpicture}}
    \caption{A distributed observer consisting of 5 nodes: each local observer $\mathcal O_i$ has available local inputs and measurements $u_i$ and $y_i$.
    Furthermore, neighbouring estimates are exchanged over a undirected communication network (dashed line).}
    \label{fig:network}
\end{figure}

To provide a visual example of the proposed architecture, in Fig. \ref{fig:network}, an undirected  network of  distributed observers with 5 nodes is shown, where the local information of each node includes the local output measurement vector $y_i$ and the local known control input vector $u_i$.

We can finally characterize the distributed estimation problem. Let $\hat x_i$ denote the estimate of $x$ produced by the local observer $\mathcal O_i$, then we define the estimation error as 
\begin{equation}
    \label{eq:error}
    e_i = x - \hat x_i.
\end{equation}
A DUIO is hence defined as follows.

\begin{definition}\label{def:dUIO}
    The set of observers $\{ \mathcal O_i \}_{i \in \mathbf N}$ is a DUIO for system \eqref{eq:system} if for all $i \in \mathbf N,$
    \begin{equation*}
        \lim_{t\rightarrow +\infty} |e_i(t)| = 0,
    \end{equation*}
    for all locally unknown inputs $\acute w_i$. \hfill$\triangleleft$
\end{definition}

That is to say a distributed observer is a DUIO if the local estimation error terms are decoupled from the disturbances and the input components that are not locally available.

\section{Distributed Unknown Input Observer Design}\label{sec:Main1}
In this section, first by assuming that the communication graph is undirected and fixed, the proposed DUIO design is presented. Then, the results are extended to scenarios when the undirected communication graph is switching over time or the communication graph is directed.

\subsection{Fundamental Results for  Undirected Networks}
The basic principle to design an unknown input observer is to derive some algebraic conditions that decouple the observer's error from the unknown disturbances/inputs \cite{chen1996design,darouach1994full,wang1975observing,guan1991novel,hou1992design,pertew2005design,bhattacharyya1978observer,commault2001unknown}.
Based on this general idea and following a pattern similar to \cite{chen1996design}, we propose the following full-order local observer $\mathcal O_i, i\in\mathbf{N}$:
\begin{equation}\label{eq:distributed observer}
\begin{split}
    \dot{z}_i &= N_i z_i + M_iB_i u_i + L_i y_i  + \chi P_i^{-1} \sum_{j=1}^N a_{ij}(\hat{x}_j - \hat x_i), \\
    \hat{x}_i &= z_i + H_i y_i,
\end{split}
\end{equation}
where $z_i \in \mathbb R^n$ is the state vector of the observer $\mathcal O_i$, matrices $N_i, M_i, H_i, P_i$ are to be designed, and $\chi$ is a real-valued design parameter. 

\begin{remark}\em
The summation in \eqref{eq:distributed observer}, although it is performed over all nodes in the network, it only includes neighbours, as $a_{ij} = 0$ if Node $i$ does not receive information from Node $j$. 
~\hfill$\triangleleft$
\end{remark}

It can be shown (see Appendix~\ref{app:error_proof}) that the estimation error of observer \eqref{eq:distributed observer} with respect to dynamics \eqref{eq:system} is given by the following differential equation:
\begin{equation}
    \label{eq:error_equation}
    \begin{aligned}
        \dot e_i &= [(I_n - H_iC_i)A - K_iC_i]e_i \\
            &\quad + (I_n - H_iC_i - M_i)B_i u_i + (I_n - H_iC_i)\bar B_i \acute w_i \\
            &\quad + [(I_n - H_iC_i)A - K_iC_i - N_i]z_i \\ 
            &\quad + [K_i + ((I_n -H_iC_i)A - K_iC_i)H_i - L_i] y_i \\
            &\quad + \chi P_i^{-1} \sum_{j = 1}^N a_{ij} (e_j - e_i).
    \end{aligned}    
\end{equation}
Now, we impose the following conditions:
\begin{subequations}
    \label{eq:uio_conditions}
    \begin{align}
        (I_n &- H_iC_i)\bar B_i = \mathbf{0}_{n\times1}, \label{eq:cond:decouple} \\
        M_i &= I_n - H_iC_i, \label{eq:cond:M} \\
        N_i &= M_i A - K_iC_i, \label{eq:cond:N} \\
        L_i &= K_i + N_iH_i. \label{eq:cond:L_i}
    \end{align}
\end{subequations}

For convenience, and as a starting point to tackle the solution of \eqref{eq:uio_conditions}, we recall the following lemma.

\begin{lemma}[\cite{chen1996design}] \label{lem:rank}
Equation \eqref{eq:cond:decouple} is solvable if and only if
\begin{equation*}
    \label{eq:rkCB}
    \rank (C_i \bar B_i) = \rank (\bar B_i),
\end{equation*}
and the general solution is given by 
\begin{equation}
    \label{eq:Hsol}
    \begin{aligned}
    H_i &= \bar B_i(C_i \bar B_i)^\dagger + Y_i [ I_{p_i} - C_i \bar B_i (C_i \bar B_i)^\dagger ] \\
        &= U_i + Y_i V_i,
    \end{aligned}
\end{equation}
where $Y_i \in \mathbb R^{n \times p_i}$ is an arbitrary matrix, and  $U_i$ and $V_i$ are defined for convenience of notation. \hfill $\square$
\end{lemma}

Lemma~\ref{lem:rank} provides a geometric condition that allows \eqref{eq:cond:decouple} in particular to be satisfied. 
If one satisfies also the other equations in the group \eqref{eq:uio_conditions}, then the estimation error in \eqref{eq:error_equation} takes the following form:
\begin{equation}
    \label{eq:uio_error}
    \dot e_i = N_i e_i + \chi P_i^{-1} \sum_{j = 1}^N a_{ij} (e_j - e_i).
\end{equation}

Before introducing the main results on the design and existence of the DUIO, we investigate the detectability properties of the system.
For convenience, we first introduce the following definition.

\begin{definition}[Extensive joint detectability]\label{def:joint_detectability}
    Let
    \begin{equation}
    \label{eq: A_i}
        A_i = (I_n - U_i C_i)A.
    \end{equation}
    System \eqref{eq:system} is extensively jointly detectable from Node $i$ if 
    \begin{equation}\label{eq:extensively jointly observable def}
        \bigcap_{i=1}^N \mathscr{U\!D}(C_i,A_i) = 0.
    \end{equation} 
    ~\hfill$\triangleleft$
\end{definition}

Like in \cite{olfati2007distributed,kamgarpour2008convergence,wang2017distributed,kim2016distributed,han2018simple,MitraAutRob:19,MitraAut:19,YangIFAC:20,Hearxiv:19,ugrinovskii2013distributed,XuIFAC:20,shen2010distributed,farina2010distributed}, we do not assume that each pair $(C_i, A)$ is observable, or even detectable, i.e., a single output measurement $y_i$ may not be sufficient in general to observe the system's state.
As remarked in \cite{wang2017distributed}, this relaxation -- that is assuming collective observability of the system -- is now consolidated in the more recent literature, although the detectability condition of Definition~\ref{def:joint_detectability} is less restrictive than the collective observability. Either collective observability or joint detectability are however global properties stemming from the cooperation of all agents.
In the following, we derive some further results on joint detectability.

By virtue of the definition of $A_i$ in \eqref{eq: A_i}, recalling \eqref{eq:Hsol}, and by inspection of \eqref{eq:cond:N}, we define
\begin{equation}\label{eq:barA_i}
    \bar A_i = (I_n - H_i C_i)A = A_i - Y_iV_iC_iA,
\end{equation}
so that we can express $N_i = \bar A_i - K_iC_i$.
Then, the convergence of the estimation errors in terms of  the detectability properties of the pair $(C_i, \bar A_i)$ will be investigated.
Accordingly, we introduce a similarity transformation matrix $T_i \in \mathbb{R}^{n\times n},i\in \mathbf{N}$,  as $T_i = \begin{bmatrix} T_{id}&T_{iu}\end{bmatrix}$ in which $T_{iu} \in \mathbb{R}^{n\times v_i}$ is an orthonormal basis of the undetectable subspace of $\left( C_i,\bar{A}_i\right)$, where  $v_i$ is the dimension of the undetectable subspace of the pair $(C_i,\bar{A}_i)$, and $T_{id}\in \mathbb{R}^{n\times (n-v_i)}$ is an orthonormal basis such that $\mathrm{Im\ }T_{id}$ is orthogonal to $\mathrm{Im\ }T_{iu}$ \cite{kim2016distributed}. 
Note that by defining $\mathscr{X} \simeq \mathbb R^n$ as the $n$-dimensional state space of the system, we have $\mathscr{X}=\mathrm{Im\ }T_{id} \oplus \mathrm{Im\ }T_{iu}$.
In the following lemmas, we investigate  such detectability properties using a geometric approach. We first prove that detectability of the pairs $(C_i, \bar A_i)$ and $(C_i, A_i)$ are equivalent, and then we provide a condition for which all the estimation errors can be steered to zero.

\begin{lemma}\label{le:equal unobservable subspace}
The undetectable subspace of the pairs $(C_i,\bar{A}_i)$ and $(C_i,A_i)$ are identical for all $Y_i \in \mathbb{R}^{n\times p_i}$.
~\hfill$\square$
\end{lemma}

\textbf{Proof}. 
By considering \eqref{eq:barA_i}, for some $F_i \in \mathbb{R}^{n\times p_i}$, one gets
\begin{equation}\label{eq:F_ifeedback}
\begin{split}
    \bar{A}_i + F_i C_i &= A_i - Y_iV_iC_i A_i + F_iC_i\\
    &= A_i + \begin{bmatrix} F_i & -Y_iV_i\end{bmatrix} \begin{bmatrix} C_i \\ C_i A_i \end{bmatrix}.
\end{split}
\end{equation}
From \eqref{eq:F_ifeedback}, it follows that
\begin{equation}\label{eq:undetect (Ci,barAi)}
    \mathscr{U\!D}\left(C_i,\bar{A}_i\right)=  \mathscr{U\!O}\left(\begin{bmatrix} C_i \\ C_i A_i \end{bmatrix},A_i\right) \bigcap \ \mathrm{Ker}\ \alpha_{A_i}^+(A_i) .
\end{equation}
Meanwhile,
\begin{equation}\label{eq:undetect (Ci,Ai)}
    \mathscr{U\!D}\left(C_i,A_i\right)=  \mathscr{U\!O}(C_i,A_i)\ \bigcap \ \mathrm{Ker}\ \alpha_{A_i}^+(A_i).
\end{equation}
By comparing \eqref{eq:undetect (Ci,barAi)} and \eqref{eq:undetect (Ci,Ai)}, to show that $\mathscr{U\!D}\left(C_i,\bar{A}_i\right)=\mathscr{U\!D}\left(C_i,A_i\right)$, we can show that the unobservable  subspace of $\left(\begin{bmatrix} C_i \\ C_i A_i \end{bmatrix},A_i\right)$ and $(C_i,A_i)$ are identical. 
It can be said that
\begin{equation}\label{eq: unobs1}
  \mathscr{U\!O}\left(\begin{bmatrix} C_i \\ C_i A_i \end{bmatrix},A_i\right)=  \bigcap_{k=1}^n \mathrm{Ker}\begin{bmatrix} C_i \\ C_i A_i \end{bmatrix} A_i^{k-1}.
\end{equation}
One can observe that
\begin{equation*}
\mathrm{Ker}\begin{bmatrix} C_i \\ C_i A_i \end{bmatrix}A_i^{k-1} = \mathrm{Ker}\ C_i A_i^{k-1} \bigcap\ \mathrm{Ker}\ C_i A_i^k,
\end{equation*}
which implies that
\begin{equation}\label{eq: unobsequal2}
    \bigcap_{k=1}^n \mathrm{Ker}\begin{bmatrix} C_i \\ C_i A_i \end{bmatrix} A_i^{k-1} = \bigcap_{k=1}^{n+1} \mathrm{Ker}\ C_i A_i^{k-1}.
\end{equation}
Moreover, the unobservable subspace of the pair $(C_i,A_i)$ is 
\begin{equation}\label{eq: unobs2}
  \mathscr{U\!O}\left(C_i,A_i\right)=  \bigcap_{k=1}^{n} \mathrm{Ker}\ C_i A_i^{k-1}.
\end{equation}
Now, from \eqref{eq: unobs1}, \eqref{eq: unobsequal2}, and \eqref{eq: unobs2}, it follows that 
\begin{equation*}\label{}
  \mathscr{U\!O}\left(C_i,A_i\right)=    \mathscr{U\!O}\left(\begin{bmatrix} C_i \\ C_i A_i \end{bmatrix},A_i\right).
\end{equation*}
Thus, from \eqref{eq:undetect (Ci,barAi)} and \eqref{eq:undetect (Ci,Ai)}, we have
\begin{equation*}\label{}
  \mathscr{U\!D}\left(C_i,A_i\right)=    \mathscr{U\!D}\left(C_i,\bar{A}_i\right),
\end{equation*}
which completes the proof.
\hfill $\blacksquare$

\begin{lemma}\label{le:T_d}
Let the system \eqref{eq:system} be extensively jointly detectable.
Then, by letting 
\begin{equation*}
 T_d=\begin{bmatrix} T_{1d}&T_{2d}&\dots&T_{Nd}\end{bmatrix},
\end{equation*}
we have
\begin{equation*}
    \mathrm{Im} \ T_d = \mathscr{X} .
\end{equation*}
~\hfill$\square$
\end{lemma}

\textbf{Proof}. 
Since $T_{iu}$ is a orthonormal basis of the undetectable subspace of $\left( C_i,\bar{A}_i\right)$, we have
\begin{equation}\label{PrepImTu}
     \mathrm{Im} \ T_{iu} = \mathscr{U\!D}(C_i,\bar{A}_i).
\end{equation}
Moreover, since  $T_{id}$ is the orthonormal basis such that $T_i$ has full rank, one gets 
\begin{equation}\label{PrepImTu2}
     \mathrm{Im} \ T_{iu} = \left( \mathrm{Im} \ T_{id}\right)^\perp.
\end{equation}
According to the definition of  $T_d$, it can be said that \cite[Chap. 0.12]{wonham1985linear}
\begin{equation*}
\begin{split}
        \mathrm{Im}\ T_d =\left( \left( \sum_{i=1}^N \mathrm{Im}\  T_{id}\right)^\perp\right)^\perp
\end{split}
\end{equation*}
implying that
\begin{equation}\label{ImTd}
\begin{split}
        \mathrm{Im}\ T_d =
         \left( \bigcap_{i=1}^N \left( \mathrm{Im}\  T_{id}\right)^\perp\right)^\perp.
\end{split}
\end{equation}
From \eqref{PrepImTu}, \eqref{PrepImTu2}, and \eqref{ImTd}, one gets
\begin{equation}\label{eq:ImTop}
\begin{split}
        \mathrm{Im}\ T_d = \left( \bigcap_{i=1}^N  \mathscr{U\!D}(C_i,\bar{A}_i)  \right)^\perp .
\end{split}
\end{equation}
From Lemma \ref{le:equal unobservable subspace}, we have
\begin{equation*}\label{}
 \mathscr{U\!D}\left(C_i,\bar{A}_i\right)= \mathscr{U\!D}\left(C_i,A_i\right).
\end{equation*}
Hence from \eqref{eq:ImTop}, we have
\begin{equation}\label{eq:ImTop2}
    \mathrm{Im}\ T_d = \left( \bigcap_{i=1}^N \mathscr{U\!D}(C_i,A_i)  \right)^\perp .
\end{equation}
Finally, by \eqref{eq:ImTop2} and under the hypothesis of extensive joint detectability \eqref{eq:extensively jointly observable def}, we have $\mathrm{Im}\ T_d=0^\perp =\mathscr{X}$ \cite[Chap.~0.12]{wonham1985linear}, which completes the proof.
\hfill $\blacksquare$

The presented lemmas let us investigate stability of the estimation errors, with the hypotheses that a solution to \eqref{eq:uio_conditions} exists.
In this case, we leverage standard Lyapunov arguments to obtain an LMI condition that guarantees stability of the (collective) error $e$, defined as the stacked vector of local observers' errors as follows:
\begin{equation*}
        e = \begin{bmatrix} e_1 ^\top& e_2 ^\top & \dots & e_N ^\top\end{bmatrix}^\top.
\end{equation*}
The stability of the proposed distributed estimation scheme is studied in the following theorem.
In this regard, the following is assumed.

\begin{assumption}
    \label{as:connected} 
    The network communication graph is connected. \hfill $\triangleleft$
\end{assumption}

\begin{theorem}[Stability]\label{th:stability} Consider the DUIO described in \eqref{eq:distributed observer} under Assumption  \ref{as:connected} and the conditions \eqref{eq:uio_conditions}.
By letting
    \begin{equation}\label{eq:Lambda_i def}
    \begin{split}
        \Lambda_i =& A^\top (I_n - C_i^\top U_i^\top) P_i + P_i(I_n - U_iC_i)A \\
        &- A^\top C_i^\top V_i^\top \bar{Y}_i^\top - \bar{Y}_i V_i C_i A \\&- C_i^\top \bar{K}_i^\top - \bar{K}_iC_i,
    \end{split}
    \end{equation}
    the estimation error $e$ converges to zero if the matrices $P_i \succ 0$, $Y_i = P_i^{-1} \bar{Y}_i$, and $K_i = P_i^{-1} \bar{K}_i$ are a feasible solution of the following LMI
    \begin{equation}\label{eq:sum Lambda_i<0}
        \sum_{i=1}^N \Lambda_i \prec 0,
    \end{equation}
    and the gain $\chi$ satisfies
    \begin{equation}\label{eq:chi def}
        \chi > \frac{\left|\Lambda + \Lambda_P^\top \left(\sum_{i=1}^N \Lambda_i \right)^{-1}\Lambda_P \right |}{2\lambda_2(\mathcal{L})},
    \end{equation}
    where 
    \begin{equation}\label{eq:Lambda Lambda_P def}
    \begin{split}
        \Lambda &= \mathrm{diag}(\Lambda_1,\Lambda_2,\ldots,\Lambda_N),\\
        \Lambda_P &= \begin{bmatrix}
        \Lambda_1 & \Lambda_2 & \ldots & \Lambda_N
        \end{bmatrix}.
    \end{split}
    \end{equation}
    ~\hfill$\square$
\end{theorem}

\textbf{Proof}. 
We show that along \eqref{eq:uio_error}, the estimation errors converge to zero. 
Accordingly, we consider the following Lyapunov candidate of the estimation errors:
\begin{equation}\label{eq:Lyapunov_func}
    V  = \sum_{i=1}^N e_i ^\top P_i e_i,
\end{equation}
which is a positive definite function of the estimation errors.
The time derivative of $V $ along \eqref{eq:uio_error} can be stated as follows:
\begin{equation}\label{eq:Vdot0}
    \dot{V} = e^\top \diag_{i \in \mathbf{N}}(N_i^\top P_i+P_iN_i)e - 2 \chi e^\top (\mathcal{L} \otimes I_n)e.
\end{equation}
Based on the conditions on $M_i$ and $N_i$ in \eqref{eq:cond:M} and \eqref{eq:cond:N}, it follows that
\begin{equation}\label{NiPi}
    \begin{aligned}
        N_i^\top P_i + P_i N_i=& \Big((I_n-H_iC_i)A - K_iC_i\Big)^\top P_i \\ &+ P_i\Big((I_n-H_iC_i)A - K_iC_i\Big).
    \end{aligned}
\end{equation}
According to  \eqref{NiPi} and the definition of $H_i$ in \eqref{eq:Hsol}, one gets
\begin{equation*}
    \begin{aligned}
        &\hspace{3ex}N_i^\top P_i + P_i N_i =\\
        &A^\top P_i + P_iA - A^\top C_i^\top U_i^\top P_i - P_i U_i C_i A \\ 
        &- A^\top C_i^\top V_i^\top Y_i^\top P_i - P_i Y_i V_i C_i A
    - C_i^\top K_i^\top P_i - P_iK_iC_i,
    \end{aligned}
\end{equation*}
which by considering \eqref{eq:Lambda_i def} with $\bar Y_i = P_iY_i $ and $\bar K_i = P_iK_i $, can be rewritten as
\begin{equation}\label{eq:Lambda_i2}
    N_i^\top P_i + P_i N_i=\Lambda_i.
\end{equation}
Now, from \eqref{eq:Lambda_i2}, \eqref{eq:Vdot0} can be restated as follows:
\begin{equation}\label{eq:Vdot1}
    \dot{V} = e^\top \Lambda e - 2 \chi e^\top (\mathcal{L} \otimes I_n)e,
\end{equation}
where $\Lambda$ is defined in \eqref{eq:Lambda Lambda_P def}. 
To analyze \eqref{eq:Vdot1}, we decompose the  error space to two subspaces. 
By defining the error space as ${\mathscr{E}\simeq \mathbb{R}^{Nn}}$, one of these subspaces is denoted by ${\mathscr{E}_c \subseteq \mathscr{E}}$ (${\dim \left(  \mathscr{E}_c\right) = n}$) which is the kernel of $\mathcal{L}\otimes I_n$ and has the form of $\mathbf{1}_N \otimes \omega, \omega \in \mathbb{R}^{n}$. 
Accordingly, the other subspace is the orthogonal complement subspace of $\mathscr{E}_c$ which is denoted by $\mathscr{E}_r \subseteq \mathscr{E}$ (${\dim \left(  \mathscr{E}_c\right) = Nn-n}$) such that ${\mathscr{E}_c \oplus \mathscr{E}_r = \mathscr{E}}$.
Thus, by considering the elements of the subspaces $\mathscr{E}_c$ and $\mathscr{E}_r$ as $e_c$ and $e_r$ ($e_c \in \mathscr{E}_c$ and $e_r \in \mathscr{E}_r$), \eqref{eq:Vdot1} yields
\begin{equation*}
    \dot{V} = e_c^\top \Lambda e_c + 2e_r^\top \Lambda e_c + e_r^\top (\Lambda - 2\chi( \mathcal{L} \otimes I_n)) e_r,
\end{equation*}
which since $e_c=\mathbf{1}_N\otimes \omega$ can be restated as follows:
\begin{equation}\label{eq:Vdot decomposed}\begin{split}
    \dot{V} =& \omega^\top \Big(\sum_{i=1}^N\Lambda_i\Big) \omega + 2e_r^\top \Lambda_P^\top \omega \\&+ e_r^\top (\Lambda - 2\chi( \mathcal{L} \otimes I_n)) e_r.
    \end{split}
\end{equation}
Moreover, as $e_r$ is orthogonal to the kernel of $\mathcal{L} \otimes I_n$, one gets \cite{OlfatiTAC:04}
\begin{equation}\label{lambda2}
-e_r ^\top (\mathcal{L} \otimes I_n) e_r \leq -\lambda_2(\mathcal{L}) e_r ^\top  e_r.
\end{equation}
Since the graph is connected from Assumption~\ref{as:connected}, $\lambda_2(\mathcal{L}) \in \mathbb{R}_{>0}$.
By considering \eqref{eq:Vdot decomposed} and \eqref{lambda2}, we have
\begin{equation}\label{eq:vdotfinal}
\begin{split}
   &\dot{V}\leq 
   - \begin{bmatrix} \omega \\e_r\end{bmatrix}^\top 
   \begin{bmatrix} -\sum_{i=1}^N \Lambda_i &-\Lambda_P  \\ -\Lambda_P^\top & ~~2\chi \lambda_2(\mathcal{L})I_{Nn}-\Lambda \end{bmatrix} 
   \begin{bmatrix} \omega\\e_r\end{bmatrix}.
   \end{split}
\end{equation}
From the inequality \eqref{eq:chi def}, one gets:
\begin{equation}\label{eq:chi def2}
    2\chi \lambda_2(\mathcal{L})I_{Nn}-\Lambda - \Lambda_P^\top \left(\sum_{i=1}^N \Lambda_i\right)^{-1}\Lambda_P \succ 0.
\end{equation}
Finally, according to \eqref{eq:chi def2} and by invoking the Schur Complement \cite{Boyd}, the negative definiteness of $\dot{V}$ in \eqref{eq:vdotfinal} can be concluded.
Thus, $V$ asymptotically converges to zero, which implies that the estimation error $e$ (and therefore all its components ${e_i,\forall i \in \mathbf{N}}$) converges to zero.
\hfill $\blacksquare$

\begin{remark}\em
From \eqref{eq:Vdot1}, it follows that
\begin{equation*}
    \dot{V} = - e^\top \left(2\chi (\mathcal{L}\otimes I_n) - \Lambda \right) e,
\end{equation*} 
where according to the proof of Theorem \ref{th:stability}, $2\chi(\mathcal{L}\otimes I_n) - \Lambda \succ 0$.
Now, by considering \eqref{eq:Lyapunov_func}, one gets
\begin{equation*}
    \dot{V} \leq - \mu V,
\end{equation*} 
where
\begin{equation}\label{eq:mu}
   \mu=\frac{\lambda_{\min}\big(2\chi (\mathcal{L}\otimes I_n) - \Lambda\big)}{\max\limits_{i\in \mathbf{N}} \big( \lambda_{\max} (P_i)\big)}.
\end{equation}
From the comparison theorem for scalar ordinary differential equations \cite[Chap.~3]{khalil2002nonlinear}, one obtains $V\leq v$ for all $t\geq 0$, where $v$ is given by
\[
v(t) = e^{-\mu t}v(0),
\]
namely $v$ converges to zero with time constant $1/{\mu}$.  
\hfill $\triangleleft$
\end{remark}

\begin{theorem}[Feasibility]\label{th:feasibility}
    If system \eqref{eq:system} is extensively jointly detectable, then the LMI \eqref{eq:sum Lambda_i<0} is always feasible for some $P_i$, $\bar Y_i = P_iY_i $, and $\bar K_i = P_iK_i $.
    ~\hfill$\square$
\end{theorem}

\textbf{Proof}. 
From \eqref{eq:Lambda_i def} and \eqref{eq:barA_i}, $\Lambda_i$ can be written as 
\begin{equation}\label{eq:Lambda_i 3}
    \Lambda_i = (\bar{A}_i - K_iC_i)^\top P_i + P_i(\bar{A}_i - K_iC_i).
\end{equation}
By considering the similarity transformation matrix 
$T_i \in \mathbb{R}^{n\times n}$, one can observe that  \cite{kim2016distributed}
\begin{equation}\label{eq:T_i property}
\begin{split}
        T_i^\top \bar{A}_i T_i =&\; \begin{bmatrix} \bar{A}_{id}&\mathbf{0}_{(n-v_i)\times v_i}\\\bar{A}_{ir}&\bar{A}_{iu}\end{bmatrix}, \\
        C_i T_i =&\; \begin{bmatrix} C_{id}&\mathbf{0}_{p_i\times v_i}\end{bmatrix},
\end{split}
\end{equation}
where the pair $(C_{id},\bar{A}_{id})$ is detectable. Based on the aforementioned formulation, without loss of generality, let the observer gains $K_i \in \mathbb{R}^{n\times p_i}$ and $P_i \in \mathbb{R}^{n\times n}$ be as follows:
\begin{equation}\label{eq:K_i and P_i def}
\begin{split}
    K_i &= T_i \begin{bmatrix} K_{id} \\ \mathbf{0}_{v_i\times p_i}\end{bmatrix}, \\
    P_i &= T_i \begin{bmatrix} P_{id}&\mathbf{0}_{(n-v_i)\times v_i} \\ \mathbf{0}_{v_i\times (n-v_i)}& P_{iu} \end{bmatrix}T_i^\top,
\end{split}
\end{equation}
where ${K_{id}\in \mathbb{R}^{(n-v_i)\times p_i}}$, ${P_{id}\in \mathbb{R}^{(n-v_i)\times (n-v_i)}\succ 0}$, and ${P_{iu}\in \mathbb{R}^{v_i\times v_i}\succ 0}$.
From the definition of $\Lambda_i$ in \eqref{eq:Lambda_i 3}, the definition of $K_i$ and $P_i$ in \eqref{eq:K_i and P_i def}, and the  decomposition performed in \eqref{eq:T_i property}, we have
\begin{equation}\label{eq:Lambda_i 2}
    \Lambda_i = T_i \begin{bmatrix} \Lambda_{id} & \Lambda_{ir}^\top \\ \Lambda_{ir} & \Lambda_{iu} \end{bmatrix} T_i^\top,
\end{equation}
where ${\Lambda_{id}\in\mathbb{R}^{(n-v_i)\times (n-v_i)}}$, ${\Lambda_{ir}\in\mathbb{R}^{v_i\times (n-v_i)}}$, and ${\Lambda_{iu}\in \mathbb{R}^{v_i\times v_i}}$ are as follows:
\begin{equation*}\label{}
\begin{split}
    &\Lambda_{id} = \Gamma_{id}^\top P_{id} + P_{id}\Gamma_{id}, \\
    &\Lambda_{ir} = P_{iu}\bar{A}_{ir},\\
&\Lambda_{iu}= \bar{A}_{iu}^\top P_{iu} + P_{iu}\bar{A}_{iu},
\end{split}
\end{equation*}
in which ${\Gamma_{id}=\bar{A}_{id}- K_{id}C_{id}}$.  
Since $T_i = \begin{bmatrix}T_{id}&T_{iu}\end{bmatrix}$, one gets:
\begin{equation}\label{eq:TDT}
\begin{split}
       T_i &\begin{bmatrix} \Lambda_{id} & \Lambda_{ir}^\top \\ \Lambda_{ir} & \Lambda_{iu} \end{bmatrix} T_i^\top
       = T_{id}\Lambda_{id}T_{id}^\top +\\
       &\quad+ T_{iu}\Lambda_{ir}T_{id}^\top + T_{id}\Lambda_{ir}^\top T_{iu}^\top + T_{iu} \Lambda_{iu} T_{iu}^\top.
\end{split}
\end{equation}
Now, from \eqref{eq:TDT}, and by defining
\begin{equation*}
\begin{split}
T_d&=\begin{bmatrix} T_{1d}&T_{2d}&\dots&T_{Nd}\end{bmatrix},\\
    \Lambda_d &= \mathrm{diag}(\Lambda_{1d},\Lambda_{2d},\dots,\Lambda_{Nd}),
    \end{split}
\end{equation*}
it follows that
\begin{equation}\label{eq:sum Lambda_i long}
\begin{split}
    &\sum_{i=1}^N \left(T_i \begin{bmatrix} \Lambda_{id} & \Lambda_{ir}^\top \\ \Lambda_{ir} & \Lambda_{iu} \end{bmatrix} T_i^\top\right) = T_d \Lambda_d  T_d^\top + \\&\quad  \sum_{i=1}^{N} \Big(T_{iu}\Lambda_{ir}T_{id}^\top  + T_{id}\Lambda_{ir}^\top T_{iu}^\top + T_{iu}\Lambda_{iu}T_{iu}^\top \Big).
    \end{split}
\end{equation}
If the system is extensively jointly detectable it follows by Lemma~\ref{le:T_d}
that $\mathrm{rank}(T_d)=n$. Hence, according to \eqref{eq:Lambda_i 2} and \eqref{eq:sum Lambda_i long}, we have
\begin{equation}\label{eq:sum Lambda_i}
    \sum_{i=1}^N \Lambda_i = T_d \left(\Lambda_d + S \right) T_d^\top,
\end{equation}
where
\begin{equation*}
    S = T_d^\dagger \! \sum_{i=1}^{N}\! \Big(\! T_{iu}\Lambda_{ir}T_{id}^\top + T_{id}\Lambda_{ir}^\top  T_{iu}^\top + T_{iu}\Lambda_{iu}T_{iu}^\top \! \Big) T_d^{\dagger\top}.
\end{equation*}
Now, according to \eqref{eq:sum Lambda_i}, since $T_d$ is row independent, the LMI~\eqref{eq:sum Lambda_i<0} is feasible if the following inequality has solution:
\begin{equation}\label{eq:LMI3}
    \Lambda_d + S \prec 0.
\end{equation}
Let us recall that $\Lambda_d = \mathrm{diag}(\Lambda_{1d},\Lambda_{2d},\dots,\Lambda_{Nd})$, where $\Lambda_{id} = \Gamma_{id}^\top P_{id} + P_{id}\Gamma_{id}$ and ${\Gamma_{id}=\bar{A}_{id} - K_{id}C_{id}}$.
Because of the detectability of the pair $(C_{id},\bar{A}_{id})$, there exists $K_{id}$ such that $\Gamma_{id}$ is Hurwitz. In this condition, according to the Lyapunov stability criterion \cite[Chap.~6]{antsaklis2006linear}, for each $\beta \in \mathbb{R}_{>0}$ there exists $P_{id}\succ 0$ such that $\Lambda_{id} = \Gamma_{id}^\top P_{id} + P_{id}\Gamma_{id}=-\beta I_{n-v_i}$. On the other hand, there exists a large enough $\beta$ such that \eqref{eq:LMI3} has solution, which guarantees the feasibility of the LMI \eqref{eq:sum Lambda_i<0}. Hence, by selecting $P_{iu}\succ 0$ and $Y_i$ arbitrarily, and  according to the definition of $K_i$ and $P_i$ in \eqref{eq:K_i and P_i def}, the LMI \eqref{eq:sum Lambda_i<0} always has solutions for $P_i$, $\bar{Y}_i$, and $\bar{K}_i$.
\hfill $\blacksquare$

Theorems~\ref{th:stability} and~\ref{th:feasibility} give \emph{constructive} sufficient conditions that can be effectively used to compute the design parameters that achieve error convergence to zero. In the next theorem, we provide necessary and sufficient existence conditions for the proposed observer to be a DUIO in the sense of Definition~\ref{def:dUIO}.

\begin{theorem}[Existence]\label{th:existence}
    The observer $\mathcal O = \{ \mathcal O_i \}_{i \in \mathbf N}$ comprising local observers in the form of \eqref{eq:distributed observer} is a DUIO for the LTI system \eqref{eq:system} if and only if the following conditions hold:
    \begin{itemize}
        \item [(i)] $\rank (C_i \bar B_i) = \rank (\bar B_i), \forall i \in \mathbf N$,
        \item [(ii)] $\displaystyle \bigcap_{i=1}^N \mathscr{U\!D}(C_i,A_i) = 0$.     ~\hfill$\square$
    \end{itemize}
 
\end{theorem}

\textbf{Proof}. 
\textit{(Sufficiency)} -- 
If (i) holds, \eqref{eq:cond:decouple} is solvable as stated in Lemma~\ref{lem:rank}.
If (ii) is true, then by Theorem~\ref{th:feasibility} we conclude that the LMI \eqref{eq:sum Lambda_i<0} admits a solution.
Therefore, we can also apply Theorem~\ref{th:stability} and conclude that such solution renders $e$ asymptotically stable, i.e., $\forall i \in \mathbf N$,
$$
    \lim_{t \rightarrow +\infty} | e_i | = 0.
$$
Therefore, $\mathcal O$ is a DUIO for \eqref{eq:system}, according to Definition~\ref{def:dUIO}.

\textit{(Necessity)} -- 
Assume now that $\mathcal O=\{ O_i \}_{i \in \mathbf N}$ is a DUIO for \eqref{eq:system}, i.e., $\forall i \in \mathbf N, \lim_{t \rightarrow +\infty} | e_i(t) | = 0$.
This immediately implies that \eqref{eq:cond:decouple} is solvable. Hence, according to Lemma~\ref{lem:rank},
(i) holds.
To prove the necessity of (ii), we proceed by contradiction and assume that there exists a nontrivial subspace $\mathscr S \subset \mathscr X$ such that
$$
    \bigcap_{i=1}^N \mathscr{U\!D}(C_i,A_i) = \mathscr S \neq 0 ,
$$
which according to \eqref{eq:undetect (Ci,Ai)} is equivalent to
\begin{equation}
    \label{eq:intersect_rearranged}
    \mathscr S = \left(\bigcap_{i=1}^{N}  \mathscr{U\!O}(C_i,A_i) \right)  \bigcap  \left( \bigcap_{i=1}^{N} \mathrm{Ker}\ \alpha_{A_{i}}^+(A_{i}) \right).
\end{equation}
Without loss of generality and for convenience of notation, we can consider eigenvalues with unit multiplicity, so that the factorization of $\alpha^+_{A_i}(s)$ is
\begin{equation}
    \label{eq:alpha_factorization}
    \alpha^+_{A_i}(s) = \underbrace{(s - \mu_{i,1}^+)}_{\alpha_{A_i,1}^+(s)} \cdots \underbrace{(s - \mu_{i,q_i}^+)}_{\alpha_{A_i,q_i}^+(s)},
\end{equation}    
for some positive $q_i \leq n$, where $\mu_{i,k} \in \mathbb C^+$ for $k = 1,\dots,q_i$.
An analogous factorization exists for $\alpha_{A_i}^-(s)$.
By the primary decomposition theorem for linear transformations \cite[Chap.~2.2.M]{antsaklis2006linear},
$\mathscr X$ is decomposed into linearly independent subspaces as
\begin{equation}
    \mathscr X \simeq \mathscr X_i^- \oplus \mathscr X_i^+ = \mathscr X_i^- \oplus \mathscr X^+_{i,1} \oplus \cdots \oplus \mathscr X^+_{i,q_i}, 
\end{equation}
where $\mathscr X_i^- = \mathrm{Ker}\ \alpha_{A_i}^-(A_i)$ and $\mathscr X_{i,k}^+ = \mathrm{Ker}\ \alpha_{A_i,k}^+(A_i)$.
Therefore, thanks to the linear independence of the modes, we have
\begin{equation}
    \mathrm{Ker}\ \alpha_{A_i}^+(A_i) \!=\! \mathrm{Ker}\ \alpha_{A_i,1}^+(A_i) \oplus \cdots \oplus \mathrm{Ker}\ \alpha_{A_i,q_i}^+(A_i),
\end{equation}
thus the second term in the right hand side of \eqref{eq:intersect_rearranged} expands as follows:
\begin{equation*}
    \bigcap_{i=1}^{N} \mathrm{Ker}\ \alpha_{A_{i}}^+(A_{i}) = \bigcap_{i=1}^{N} \mathscr X^+_{i,1} \oplus \cdots \oplus \mathscr X^+_{i,q_i}.
\end{equation*}
Since $\mathscr S \neq 0 $, there exists $\mathscr X_\cap^+ \subseteq \mathscr X^+_{i}, \forall i \in \mathbf N,$ whose intersection with the unobservable subspaces is nontrivial.
Namely, $\mathscr X_\cap^+ \subseteq \mathscr S$ is by construction an $A_i$-invariant subspace of an undetectable mode that is common to all nodes, i.e., $\forall i \in \mathbf N$, it satisfies
\begin{equation}
    \label{eq:eigenvalue_Ai}
    \mathscr X_\cap^+\! = \!\{ x \in \mathscr X:(A_i - \mu_i I_n)x = \mathbf{0}_{n\times 1},  x \neq \mathbf{0}_{n\times 1}\}, 
\end{equation}
for some $\mu_i \in \mathbb C^+$. 
By Lemma~\ref{le:equal unobservable subspace}, Equations \eqref{eq:intersect_rearranged}--\eqref{eq:eigenvalue_Ai} hold for $\bar A_i$ as well, thus we let $v \in \mathscr X_\cap^+$ be one of such common undetectable modes, and since $\mathscr S \subseteq \mathrm{Ker}\ C_i$ for all $i \in \mathbf N$, it holds that
\begin{equation}\label{eq:barAiv}
    (\bar A_i - K_iC_i)v = \bar A_iv.
\end{equation}
By stacking the error components and from \eqref{eq:uio_error}, we obtain 
\begin{equation}
    \label{eq:uio_error_stack}
    \begin{aligned}
    \dot e &= \left( \diag_{i \in \mathbf N} (N_i) - \chi \diag_{i \in \mathbf N} \left( P_i^{-1} \right)(\mathcal L \otimes I_n) \right) e \\
        & = (\Phi - \Pi) e,
    \end{aligned}
\end{equation}
where the definitions of $\Phi$ and $\Pi$ follow trivially from the equality.
Let $\bar e = \mathbf 1_N \otimes v$. Since $\bar e \in \mathrm{Ker}\, \Pi$, it follows that $\Pi \bar e=\mathbf{0}_{Nn\times 1}$.
Moreover, for each block of $\Phi$, $\bar e$ satisfies \eqref{eq:barAiv} and the eigenvalue relation \eqref{eq:eigenvalue_Ai}. Therefore,
\begin{equation*}
    \label{eq:bare_eigen}
    \begin{aligned}
    (\Phi - \Pi) \bar e &= \diag_{i \in \mathbf N}(\bar A_i - K_iC_i) \bar e = \diag_{i \in \mathbf N}(\bar A_i)\bar e \\
    &= \begin{bmatrix}\mu_1 & \dots & \mu_N\end{bmatrix}^\top \otimes \bar e.
    \end{aligned}
\end{equation*}
Now, choosing $\bar e$ as the initial condition for \eqref{eq:uio_error_stack} produces an error dynamics along the direction of the unstable mode $v$.
This contradicts the asymptotic stability hypothesis, and therefore (ii) must be true.
\hfill $\blacksquare$

It should be noted that we have formulated Theorem~\ref{th:existence} in a way to express the similarities of the conditions derived in our approach to the classical existence conditions \cite[Theorem~1]{chen1996design} for the centralized case.
We also remark that (ii) is a necessary and sufficient condition also appearing in \cite{ugrinovskii2013distributed}.

In the following subsection, we show  how the proposed DUIO can be extended to more complex scenarios under some conditions, such as graphs with switching topologies and directed networks.

\subsection{Extension to  Switching Topologies or Directed Networks}\label{subsec:extension}

The results presented in Theorem \ref{th:stability} are based on the assumption that the communication graph is undirected and its links are steady and not failing over time. 
However, by suitably modifying $\chi$, the obtained results can be extended to more general scenarios such as distributed estimation in the presence of switching topologies and distributed estimation in directed networks.

 In the presence of switching topologies, let $\mathcal{G}(t)$ describe a communication graph switching over time.  
 Accordingly, the distributed observer proposed in \eqref{eq:distributed observer} should be modified as follows:
\begin{equation}\label{eq:distributed observer:switching}
\begin{split}
    \dot{z}_i &= N_i z_i + M_iB_i u_i + L_i y_i  + \chi P_i^{-1} \sum_{j=1}^N a_{ij}(t)(\hat{x}_j - \hat x_i), \\
    \hat{x}_i &= z_i + H_i y_i,
\end{split}
\end{equation}
where $a_{ij}(t)=1$ if there exits a communication link between Node $i$ and Node $j$ at time $t$, and it is zero otherwise.
We consider an infinite time sequence $t_0$, $t_1$, $t_2$, $\ldots$ starting at $t_0=0$, at which $\mathcal{G}(t)$ switches to $\mathcal{G}_{k},k=0,1,2,\dots$, while remaining connected. 
By considering a common Lyapunov function for the set of switching topologies the same as in \eqref{eq:Lyapunov_func} and following the same steps as in the proof of Theorem \ref{th:stability}, one gets
\begin{equation}\label{eq:vdotfina:switching}
\begin{split}
   &\dot{V}\leq 
   - \begin{bmatrix} \omega \\e_r\end{bmatrix}^\top 
   \begin{bmatrix} -\sum_{i=1}^N \Lambda_i &-\Lambda_P  \\ -\Lambda_P^\top & ~~2\chi \lambda_2(\mathcal{L}_k)I_{Nn}-\Lambda \end{bmatrix} 
   \begin{bmatrix} \omega\\e_r\end{bmatrix},
   \end{split}
\end{equation}
where $\mathcal{L}_k$ is the Laplacian matrix associated with $\mathcal{G}_k$. In this condition, according to \eqref{eq:vdotfina:switching} and 
the Schur complement, $\dot{V}$ is negative definite if 
\begin{equation}\label{eq:switching chi def}
    \chi > \frac{\left|\Lambda + \Lambda_P^\top \left(\sum_{i=1}^N \Lambda_i \right)^{-1}\Lambda_P \right |}{2\mathcal{C}(N)},
\end{equation}
where $\mathcal{C}(N)$ is a lower bound for the algebraic connectivity of  graphs with $N$ nodes which just depends on $N$ (see \cite{PrraniTAC:16} and \cite{Chung:97}). Therefore, as $\dot{V}$ is negative definite, $e$ converges to zero.

Now, let the network communication graph be fixed and strongly connected.
Assumption \ref{as:CB} implies that the Laplacian matrix associated with the communication graph is semidefinite. 
However, it is possible to modify the proposed DUIO such that by weighting the consensus terms, the obtained results in Theorem \ref{th:stability} are extendable to directed network as well. 
In this regard, we first introduce the following lemma.

\begin{lemma}[\cite{li2017cooperative}]\label{le:Laplacian} 
Assume $\mathcal{G}$ is a strongly connected directed graph. Then, there exists a unique positive row vector
$r = \begin{bmatrix} r_1&r_2&\ldots&r_N\end{bmatrix}$
such that $r\mathcal{L}=\mathbf{0}_{1\times N}$ and $r\mathbf{1}_N=N$, and by defining $R:=\mathrm{diag}(r_1,\ldots,r_N)$, the symmetric matrix $\hat{\mathcal{L}}:=R\mathcal{L}+\mathcal{L}^\top R$ is positive semidefinite. Furthermore, $\mathbf{1}_N^\top \hat{\mathcal{L}}=\mathbf{0}_{1\times N}$, $\hat{\mathcal{L}}\mathbf{1}_N=\mathbf{0}_{N\times 1}$, and $\lambda_1 = 0$ is a simple eigenvalue of $\hat{\mathcal{L}}$ while the other eigenvalues of $\hat{\mathcal{L}}$ are positive real.     ~\hfill$\square$
\end{lemma}

By weighting the consensus terms by $r_i,i\in \mathbf{N},$ the DUIO  \eqref{eq:distributed observer} can be modified as follows:
\begin{equation}\label{modifiedUIO}
    \begin{split}
    \dot{z}_i &= N_i z_i + M_iB_i u_i + L_i y_i  + \chi r_i P_i^{-1} \sum_{j=1}^N a_{ij}(\hat{x}_j - \hat x_i), \\
    \hat{x}_i &= z_i + H_i y_i.
\end{split}
\end{equation}
Considering the same Lyapunov function as in \eqref{eq:Lyapunov_func}, and following the same procedure as in the proof of Theorem~\ref{th:stability}, one gets
\begin{equation*}
\begin{split}
    \dot{V} &= \sum_{i=1}^N  e_i ^\top \Lambda_i e_i + \chi \sum_{i=1}^N r_i \Bigg( \sum_{j=1}^N  (a_{ij}+a_{ji}) e_j^\top e_i \\
    &\quad\, -2a_{ij} e_i^\top e_i \Bigg), 
\end{split}
\end{equation*}
which can be rewritten as follows:
\begin{equation}\label{eq: Vdotmodified}
\begin{split}
    \dot{V}
    &= \sum_{i=1}^N  e_i ^\top \Lambda_i e_i + \chi \sum_{i=1}^N \Bigg(\sum_{j=1}^N  (r_i a_{ij}+ r_j a_{ji}) e_j^\top e_i \\
    &\quad\, -2 r_i a_{ij} e_i^\top e_i\Bigg).
\end{split}
\end{equation}
According to the definition of $\Lambda$ and $\hat{\mathcal{L}}$, from \eqref{eq: Vdotmodified} it follows that
\begin{equation}\label{eq: Vdotmodified2} 
\begin{split}
    \dot{V} =  e^\top \Lambda e  - \chi e ^\top \left( \hat{\mathcal{L}}\otimes I_n\right) e.
\end{split}
\end{equation}
Since $\mathcal{G}$ is strongly connected, by considering Lemma~\ref{le:Laplacian}, $\hat{\mathcal{L}}$ is symmetric positive semidefinite, $\mathbf{1}_N^\top \hat{\mathcal{L}}=\mathbf{0}_{1\times N}$, $\hat{\mathcal{L}}\mathbf{1}_N=\mathbf{0}_{N\times 1}$, and $\hat{\mathcal{L}}$ has one zero eigenvalue and $N-1$ real positive eigenvalues.
By following the same procedure as in the proof of Theorem~\ref{th:stability}, from \eqref{eq: Vdotmodified2} one gets
\begin{equation*}
\begin{split}
   \dot{V}\leq - \begin{bmatrix} \omega \\e_r\end{bmatrix}^\top \begin{bmatrix} -\sum_{i=1}^N \Lambda_i &-\Lambda_P  \\ -\Lambda_P^\top & ~~2\chi \lambda_2(\hat{\mathcal{L}})I_{Nn}-\Lambda \end{bmatrix} \begin{bmatrix} \omega\\e_r\end{bmatrix},
\end{split}
\end{equation*}
which, by the Schur complement, is negative definite if 
    \begin{equation}\label{eq:directed chi def}
        \chi > \frac{\left|\Lambda + \Lambda_P^\top \left(\sum_{i=1}^N \Lambda_i \right)^{-1}\Lambda_P \right |}{2\lambda_2(\hat{\mathcal{L}})}.
    \end{equation}
Hence, as $\dot{V}$ is negative definite, $e$ converges to zero.

\section{Simulation Results}\label{sec:Sim}
    
 The accuracy of the proposed DUIO is evaluated in this section. We consider an LTI system in the form of \eqref{eq:system} where
\begingroup
\allowdisplaybreaks
\begin{align*}
    A &= \begin{bmatrix} 
        13&1&17&5&-16&2\\
        2&6&3&-1&8&4\\
        0&1&-8&-7&-16&5\\
        -2&-13&-15&-15&5&7\\
        -7&43&15&3&-11&8\\
        6&-7&1&2&1&-9\\
      \end{bmatrix}, \\
    B &= \begin{bmatrix} 
      1&0&0\\0&1&0\\0&0&1\\0&0&0\\0&0&0\\0&0&0
    \end{bmatrix},
    D = \begin{bmatrix} 0\\0\\0\\0\\0\\1 \end{bmatrix}.
\end{align*}%
\endgroup
We assume that
\begin{equation*}
\begin{split}
    B_1 &= \begin{bmatrix} 1&0&0&0&0&0 \end{bmatrix}^\top,  
     B_2 = \begin{bmatrix} 0&1&0&0&0&0 \end{bmatrix}^\top,\\ 
    B_3 &= \begin{bmatrix} 0&0&1&0&0&0 \end{bmatrix}^\top,
     B_4 = \begin{bmatrix} 0&0&0&0&0&0\end{bmatrix}^\top,
\end{split}
\end{equation*}
and the system's state vector is  as $x=[x^1~x^2~x^3~\cdots~x^6]^\top$.
Accordingly, we have
\begin{equation*}
\begin{split}
    &\bar{B}_1 = \begin{bmatrix} B_2 & B_3&D \end{bmatrix},\ \bar{B}_2 = \begin{bmatrix} B_1 & B_3&D \end{bmatrix},\\ 
    &\bar{B}_3 = \begin{bmatrix} B_1 & B_2&D \end{bmatrix}, \ \bar{B}_4 = \begin{bmatrix} B_1 & B_2&B_3&D \end{bmatrix}.
\end{split}
\end{equation*}
Moreover, the output matrices are considered as follows:
\begingroup
\allowdisplaybreaks
\begin{align*}
    C_1 &=\begin{bmatrix}               
      0&1&0&0&0&0\\
      0&0&1&0&0&0\\
      1&0&0&0&0&1 \end{bmatrix},
    C_2 =\begin{bmatrix}
      0&1&1&0&0&0\\
      1&0&1&0&0&0\\
      0&0&1&0&0&1\end{bmatrix},\\
    C_3 &=\begin{bmatrix}
      1&0&0&0&0&0\\
      0&1&0&0&0&0\\
      1&0&1&0&0&1\end{bmatrix},
    C_4 =\begin{bmatrix}
      1&0&0&1&0&1\\
      0&1&0&0&0&0\\
      0&0&1&0&0&0\\
      0&1&0&0&0&1\end{bmatrix}.
\end{align*}
\endgroup

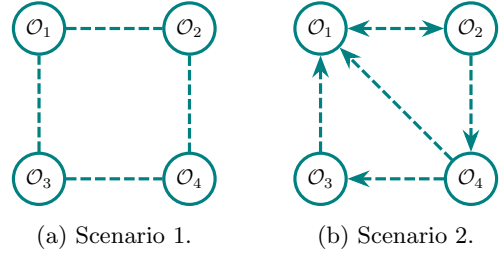
\begin{figure}[]
\centering
\subfloat[Scenario~\ref{sce:connected}.]{
    \label{fig:sce1}
    \begin{tikzpicture}
        \begin{scope}[every node/.style={scale=0.8,circle,very thick,draw,color=teal, text=black}] 
            \node (1) at (0,2) {$\mathcal O_1$};
            \node (2) at (2,2) {$\mathcal O_2$};
            \node (3) at (0,0) {$\mathcal O_3$};
            \node (4) at (2,0) {$\mathcal O_4$};
        \end{scope}
        \begin{scope}[>={Stealth[black]},
                      every edge/.style={draw=teal, very thick, dash pattern=on 4pt off 1.5pt}] 
        \path [-] (1) edge node {} (2);
        \path [-] (2) edge node {} (4);
        \path [-] (3) edge node {} (1);
        \path [-] (3) edge node {} (4);
        \end{scope}
    \end{tikzpicture}
    \label{fig:topo_connected}
}\hspace{2em}
\subfloat[Scenario~\ref{sce:directed}.]{
    \label{fig:topo_directed}
    \begin{tikzpicture}
        \begin{scope}[every node/.style={scale=0.8,circle,very thick,draw,color=teal, text=black}] 
            \node (1) at (0,2) {$\mathcal O_1$};
            \node (2) at (2,2) {$\mathcal O_2$};
            \node (3) at (0,0) {$\mathcal O_3$};
            \node (4) at (2,0) {$\mathcal O_4$};
        \end{scope}
        \begin{scope}[>={Stealth[teal]},
                      every edge/.style={draw=teal, very thick, dash pattern=on 4pt off 1.5pt}] 
        \path [<->] (1) edge node {} (2);
        \path [->] (2) edge node {} (4);
        \path [->] (4) edge node {} (1);
        \path [->] (4) edge node {} (3);
        \path [->] (3) edge node {} (1);
        \end{scope}
    \end{tikzpicture}
    
    }
    \caption{Network communication topologies in Scenarios~\ref{sce:connected} and~\ref{sce:directed}.}
    \label{fig:topo_sce1_2}
\end{figure}

\begin{figure}[]
\centering
\subfloat[Topology 1.]{
\begin{tikzpicture}
\begin{scope}[every node/.style={scale=0.8,circle,very thick,draw,color=teal, text=black}] 
    \node (1) at (0,2) {$\mathcal O_1$};
    \node (2) at (2,2) {$\mathcal O_2$};
    \node (3) at (0,0) {$\mathcal O_3$};
    \node (4) at (2,0) {$\mathcal O_4$};
\end{scope}
\begin{scope}[>={Stealth[black]},
              every edge/.style={draw=teal, very thick, dash pattern=on 4pt off 1.5pt}] 
\path [-] (1) edge node {} (2);
\path [-] (1) edge node {} (3);
\path [-] (1) edge node {} (4);
\path [-] (2) edge node {} (4);
\path [-] (3) edge node {} (4);
\end{scope}\end{tikzpicture}}\hspace{2em}
 \subfloat[Topology 2.]{
\begin{tikzpicture}
\begin{scope}[every node/.style={scale=0.8,circle,very thick,draw,color=teal, text=black}] 
    \node (1) at (0,2) {$\mathcal O_1$};
    \node (2) at (2,2) {$\mathcal O_2$};
    \node (3) at (0,0) {$\mathcal O_3$};
    \node (4) at (2,0) {$\mathcal O_4$};
\end{scope}
\begin{scope}[>={Stealth[black]},
              every edge/.style={draw=teal, very thick, dash pattern=on 4pt off 1.5pt}] 
\path [-] (1) edge node {} (2);
\path [-] (2) edge node {} (4);
\path [-] (3) edge node {} (4);
\end{scope}\end{tikzpicture}}\\
 \subfloat[Topology 3.]{
\begin{tikzpicture}
\begin{scope}[every node/.style={scale=0.8,circle,very thick,draw,color=teal, text=black}] 
    \node (1) at (0,2) {$\mathcal O_1$};
    \node (2) at (2,2) {$\mathcal O_2$};
    \node (3) at (0,0) {$\mathcal O_3$};
    \node (4) at (2,0) {$\mathcal O_4$};
\end{scope}
\begin{scope}[>={Stealth[black]},
              every edge/.style={draw=teal, very thick, dash pattern=on 4pt off 1.5pt}] 
\path [-] (1) edge node {} (2);
\path [-] (2) edge node {} (4);
\path [-] (3) edge node {} (2);
\end{scope}\end{tikzpicture}}~~~~~~~~
 \subfloat[Topology 4.]{
\begin{tikzpicture}
\begin{scope}[every node/.style={scale=0.8,circle,very thick,draw,color=teal, text=black}] 
    \node (1) at (0,2) {$\mathcal O_1$};
    \node (2) at (2,2) {$\mathcal O_2$};
    \node (3) at (0,0) {$\mathcal O_3$};
    \node (4) at (2,0) {$\mathcal O_4$};
\end{scope}
\begin{scope}[>={Stealth[black]},
              every edge/.style={draw=teal, very thick, dash pattern=on 4pt off 1.5pt}] 
\path [-] (1) edge node {} (2);
\path [-] (1) edge node {} (3);
\path [-] (4) edge node {} (3);
\end{scope}\end{tikzpicture}}\\
\caption{Switching set of the network topologies in Scenario~\ref{sce:switching}.}
\label{fig:topo_switching}
\end{figure}
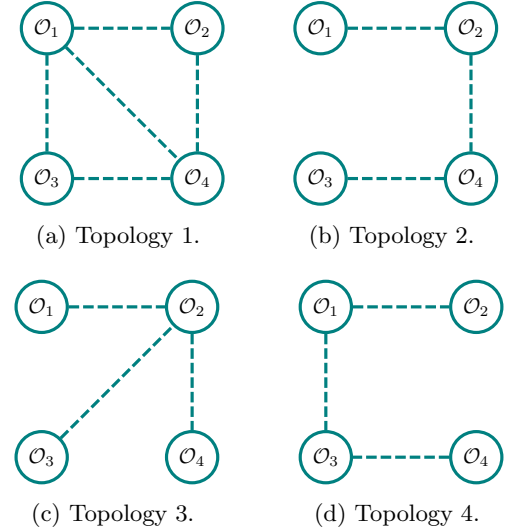

Without loss of generality, the control input is selected as $u=\begin{bmatrix} -(Fx)^\top&\vartheta\end{bmatrix}^\top$, where $F\in \mathbb{R}^{3\times 6}$ is provided in Appendix~\ref{ap:parameters} and $\vartheta$ denotes the band-limited white noise with noise power set to $1$. 
The proposed distributed estimation strategies are evaluated in three scenarios corresponding to steadily connected, strongly connected directed, and switching connected topologies, as shown in Figs. \ref{fig:topo_sce1_2}--\ref{fig:topo_switching}, respectively.


\begin{scenario}[Undirected graph]\label{sce:connected}\em
In the first scenario, the nodes are assumed to be connected via the unweighted undirected  communication  graph  depicted in Fig.~\ref{fig:topo_connected} implying that $\lambda_2(\mathcal{L})=2$. 
Distributed state estimation is based on the distributed observers \eqref{eq:distributed observer} where $N_i$, $M_i$, $L_i$, $H_i$, and $P_i$ are obtained from the solution of the LMI \eqref{eq:sum Lambda_i<0} as given in Appendix~\ref{ap:parameters}. 
It should be noted that the solution of the LMI is obtained by using the CVX toolbox \cite{BoydVandBook:04}. 
Moreover, following \eqref{eq:chi def}, $\chi$ is set to $84.81$. 
Under these conditions, the estimated state vectors of the observers along with the real state vector of the system are illustrated in Fig.~\ref{fig:states_connected}. 
According to the figure, the estimated state vectors of all the observers converge to the true state vector of the system asymptotically.
From \eqref{eq:mu}, the time constant is calculated as $4.844\times 10^{-2}$. In this regard, the evolution of the Lyapunov function $V$ along with $e^{-\mu t}V(0)$ is depicted in Fig. \ref{fig:V_connected}.

\begin{figure}
\centering
\includegraphics[width=0.47\textwidth]{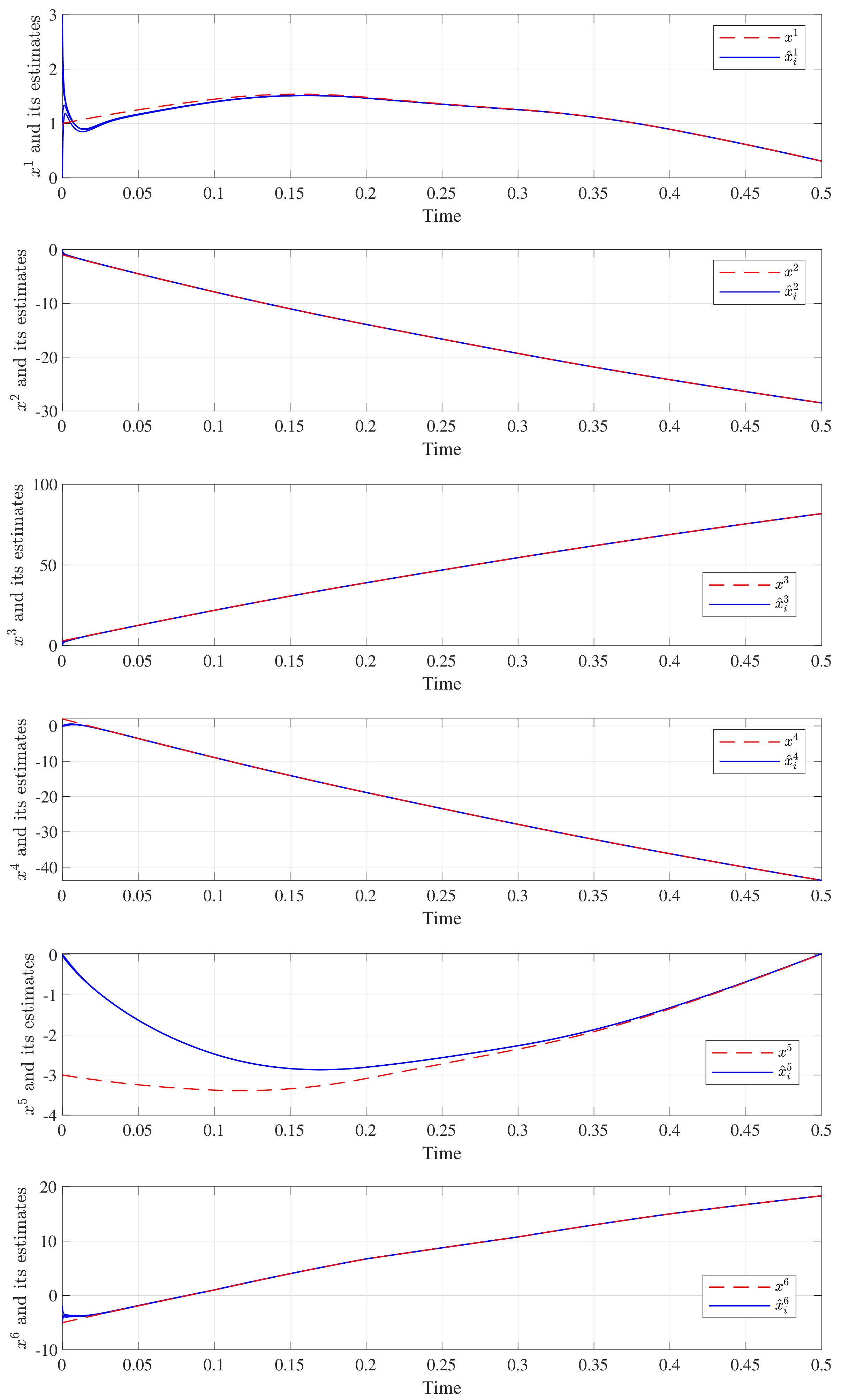}
\caption{Estimated state vectors along with the state vector of the dynamical system in the presence of unknown inputs for all the nodes in Scenario~\ref{sce:connected}.}
\label{fig:states_connected}
\end{figure}

\begin{figure}
\centering
\includegraphics[width=0.4\textwidth]{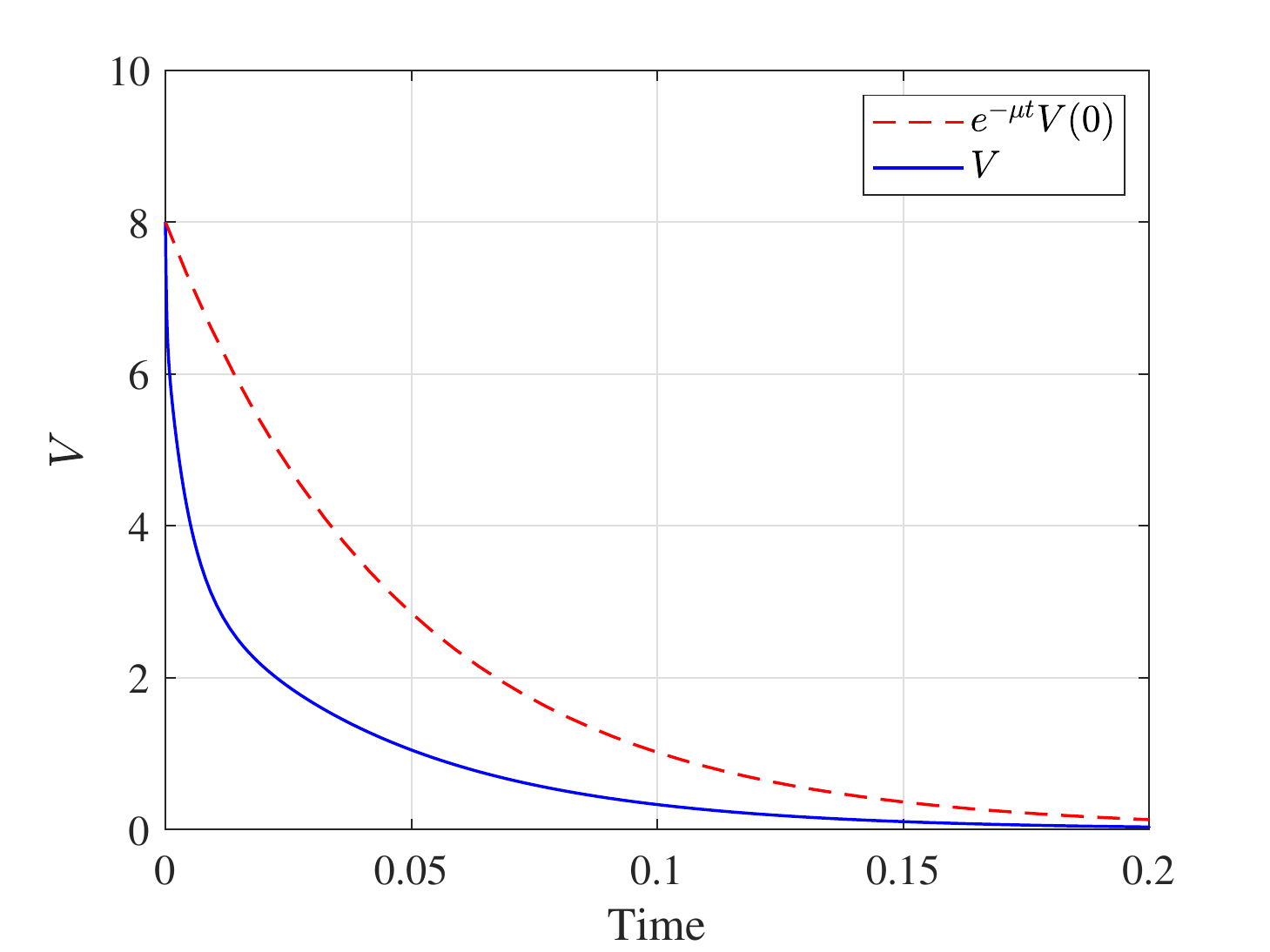}
\caption{Lyapunov function $V$ generated by the proposed distributed observer in  Scenario~\ref{sce:connected}.}
\label{fig:V_connected}
\end{figure}

\end{scenario}

\begin{scenario}[Directed graph]\label{sce:directed}\em
In the second scenario, the nodes are assumed to be connected via the unbalanced directed  communication  graph  depicted in Fig.~\ref{fig:topo_directed}. Distributed state estimation is based on the distributed observers given in \eqref{modifiedUIO} where $N_i$, $M_i$, $L_i$, $H_i$, and $P_i$ still are the same as Scenario~\ref{sce:connected} as given in Appendix~\ref{ap:parameters}. According to Lemma~\ref{le:Laplacian}, $R=\diag(0.5714,1.714,0.5714,1.143)$, and following \eqref{eq:directed chi def}, $\chi$ is set to $234.0$. 
Under these conditions, the estimated state vectors of the observers along with the true state vector of the system are illustrated in Fig.~\ref{fig:states_directed}. According to the figure, the estimated state vectors of all the observers converge to the true state vector of the system asymptotically.


\begin{figure}
\centering
\vspace*{1.1ex} 
\includegraphics[width=0.47\textwidth]{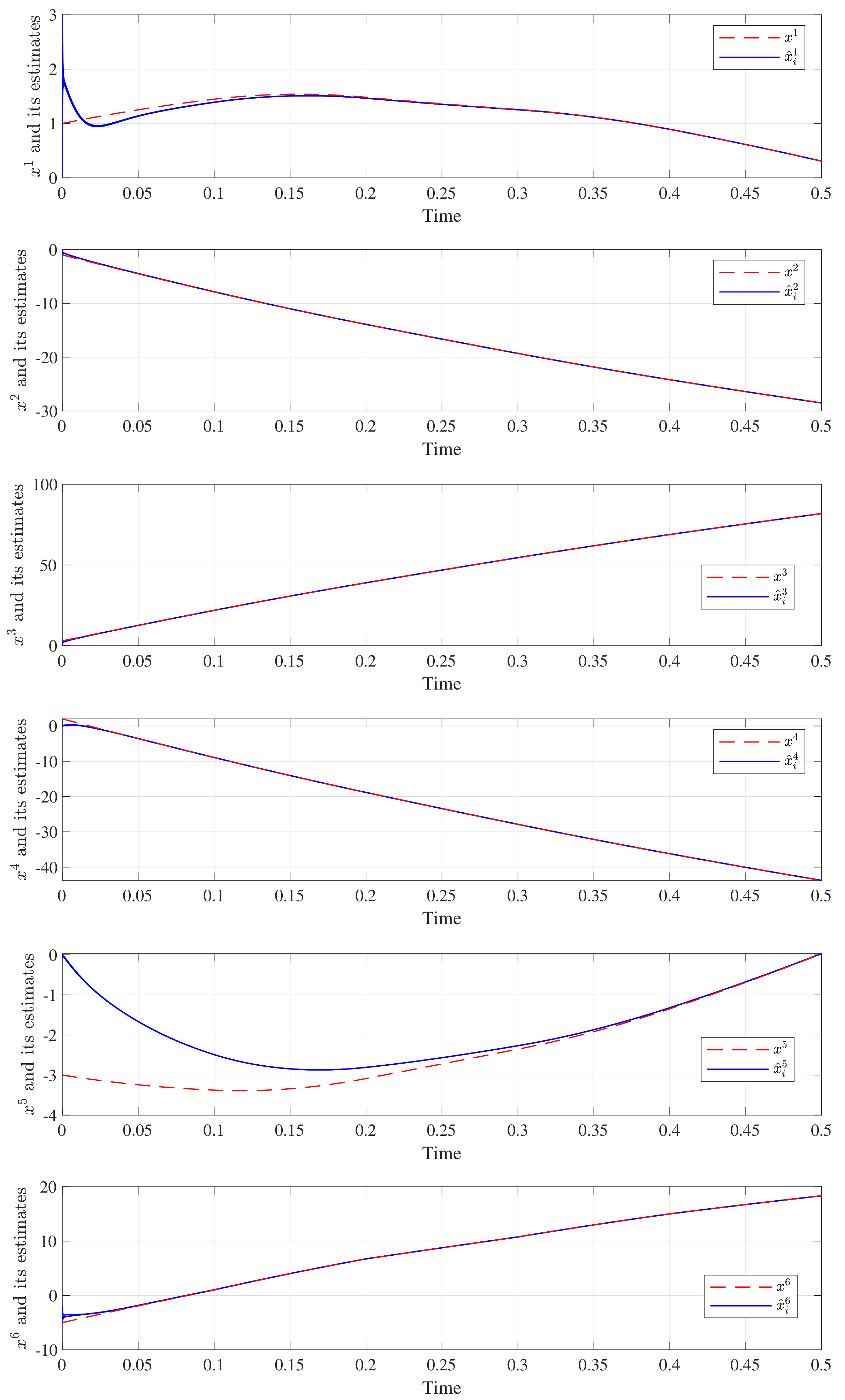}
\caption{Estimated state vectors along with the state vector of the dynamical system in the presence of unknown inputs for all the nodes  in Scenario~\ref{sce:directed}.}
\label{fig:states_directed}
\end{figure}
\end{scenario}

\begin{scenario}[Switching undirected graph]\label{sce:switching}\em
In the third scenario, the nodes are assumed to be connected under the switching communication topology depicted in Fig.~\ref{fig:topo_switching}, such that the information exchange starts from Topology 1 and switches to the next topology every $0.1$ second (after Topology 4 the graph switches back to Topology 1). 
Distributed state estimation is based on the distributed observers given in \eqref{eq:distributed observer:switching} where $N_i$, $M_i$, $L_i$, $H_i$, and $P_i$ are the same as Scenario~\ref{sce:connected} as given in Appendix~\ref{ap:parameters}. 
Moreover, $\mathcal{C}(4)$ is calculated as $4.167\times 10^{-2}$, and $\chi$ is set to $4.024\times 10^{3}$ by following \eqref{eq:switching chi def}. Under these conditions, the estimated state vectors of the observers along with true state vector of the system are illustrated in Fig.~\ref{fig:states_switching}, verifying that the estimated state vectors of all the observers converge to the true state vector of the system asymptotically.

\begin{figure}
\centering
\includegraphics[width=0.47\textwidth]{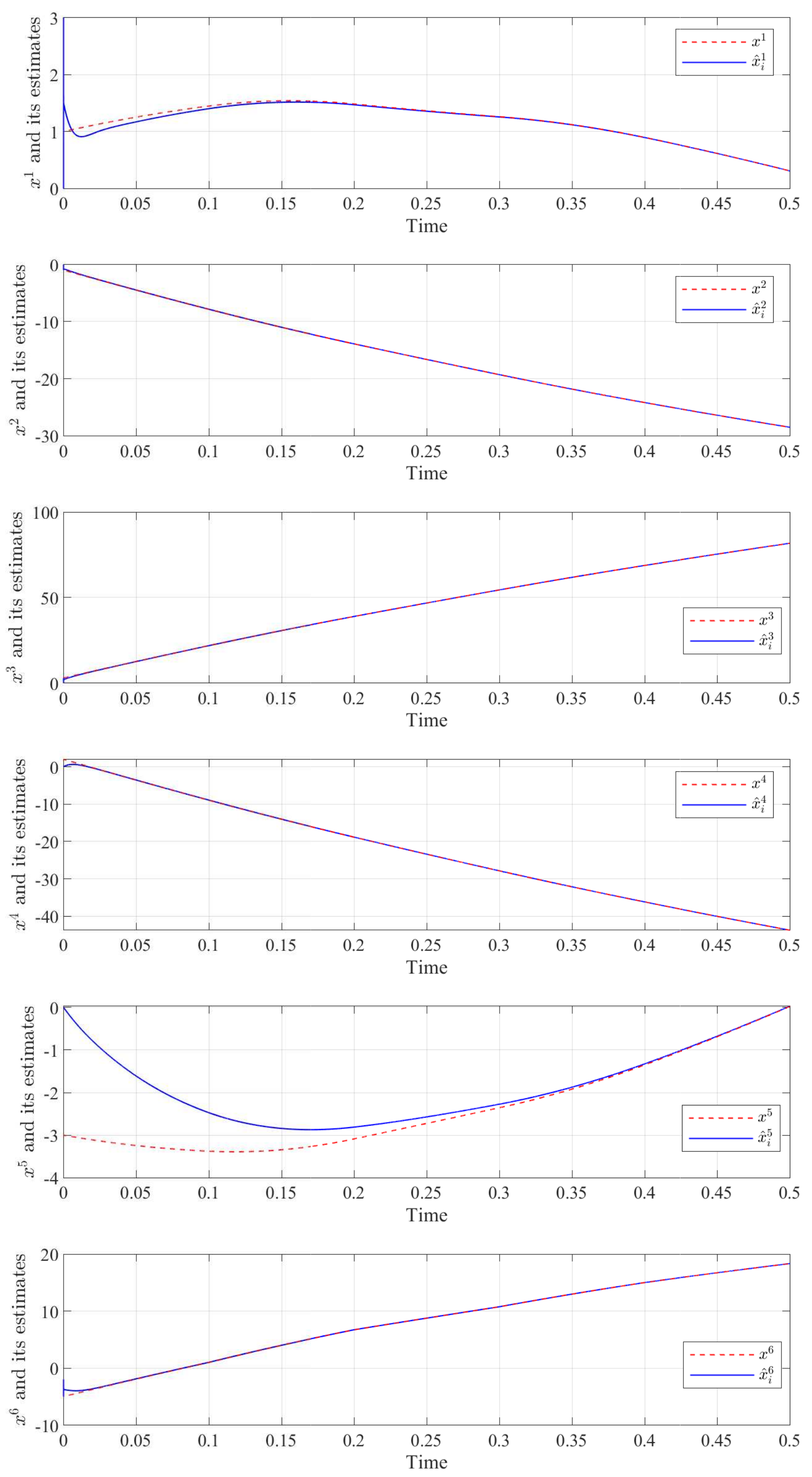}
\caption{Estimated state vectors along with the state vector of the dynamical system in the presence of unknown inputs for all the nodes in Scenario~\ref{sce:switching}.}
\label{fig:states_switching}
\end{figure}

\end{scenario}

\section{Conclusions and Future Work}\label{sec:Conc}
Distributed state estimation of a class of LTI systems was addressed, where the system outputs are measured via a network of sensors distributed within $N$ nodes, and the local measurements at each node are not sufficient for local state estimation.
We proposed a DUIO consisting of $N$ local observers co-located with the $N$ nodes and connected via a communication network such that the full state vector of the system is estimated by each local observer.
The main motivation for proposing this distributed solution is to account for partial measurements, but more remarkably for inputs that may not be available locally at a node, together with other unknown disturbances.
The feasible solution of an LMI provides adequate choices of parameters that guarantee convergence of the estimation errors, under some joint detectability conditions.
Furthermore, necessary and sufficient existence conditions are given that are in line with existing theorems for the centralized cases.
Finally, we provide modified versions of our main result allowing us to include more complex scenarios in our study, such as switching network topologies and directed communication links.
It should be noted that this study was a primary effort on DUIOs, and many problems such as designing DUIOs in the presence of measurement noise as well as expanding the obtained results to discrete-time domains remain open
to be studied as future work.

\bibliographystyle{ieeetr}
\bibliography{Automatica}

\appendix
\textbf{Appendix}
\section{Derivation of Equation \eqref{eq:error_equation}}\label{app:error_proof}
The equation to be proved is obtained by expanding the error definition \eqref{eq:error} along with the system dynamics \eqref{eq:system}, the output equation \eqref{eq:output}, and the local observer \eqref{eq:distributed observer}.
We start by noting that 
\begin{equation}
    \label{eq:app:error1}
    e_i = x - z_i - H_i y_i = (I_n - H_i C_i)x - z_i.
\end{equation}
Taking the time derivative of \eqref{eq:app:error1} yields
\begin{equation*}
\begin{aligned}
    \dot e_i =& (I_n - H_i C_i) (Ax + B_i u_i + \bar B_i \acute w_i)  \\
             & - N_i z_i - M_i B_i u_i - L_i y_i - \chi P_i^{-1}\sum_{j=1}^N a_{ij} (\hat x_j - \hat x_i),
\end{aligned}
\end{equation*}
which by adding and subtracting the term $(I_n - H_iC_i)A \hat x_i$ to the right-hand side and since $ \hat{x}_i = z_i + H_i y_i$, can be restated as follows:
\begin{equation}\label{eq:app:doterror1}
\begin{aligned}
    \dot e_i =& (I_n - H_i C_i)A e_i + (I_n - H_i C_i - M_i)B_i u_i \\
             & + (I_n - H_i C_i)\bar B_i \acute w_i + (I_n - H_i C_i)A(z_i + H_iy_i) \\
             & - N_i z_i - L_i y_i - \chi P_i^{-1}\sum_{j=1}^N a_{ij} (\hat x_j - \hat x_i).
\end{aligned}
\end{equation}
According to the definition of $e_i$, one gets
\begin{equation}\label{zeroterm1}
-K_i C_i e_i + K_i C_i(x - \hat x_i)=\mathbf 0_{n \times 1}.
\end{equation} 
Since $ \hat{x}_i = z_i + H_i y_i$, from \eqref{zeroterm1}, it follows that
\begin{equation}\label{zeroterm2}
-K_i C_i e_i + K_i y_i - K_i C_i z_i - K_i C_i H_i y_i=\mathbf 0_{n \times 1}.
\end{equation} 
We note that 
\begin{equation}\label{xixj}
\hat x_j - \hat x_i = x - \hat x_i - (x - \hat x_j) = e_i - e_j.
\end{equation}
Now, by adding the zero term \eqref{zeroterm2} to the right hand side of \eqref{eq:app:doterror1} and by considering \eqref{xixj}, after grouping similar terms, we finally obtain \eqref{eq:error_equation}.
\hfill $\blacksquare$

\section{Simulation Parameters in Section V} \label{ap:parameters}
In this section, we keep $4$ significant digits for noninteger elements in all matrices. The following parameters are used for all three scenarios.
{\footnotesize  
\allowdisplaybreaks
  \begin{align*}
    F &= \begin{bmatrix}
        7.445 &  15.70 &  24.16 &  11.19 & -19.81 &   8.128 \\
    5.254  &  4.307 &   8.581  &  6.864  &  7.416  & -2.586\\
   -4.382 & -23.23 & -33.65 & -30.91 &  -6.951  & 18.01
    \end{bmatrix}, \\
    \smallskip
    P_1 &= P_2 = P_3 = P_4 = 0.1000\times I_6 ,\\
    N_1 &= \begin{bmatrix} -205.0 &   -3.400  &  8.750 &   5.000 & -16.00& -216.0\\
    0& -214.3   & 0.4000     &    0       &  0  &  0\\
    0  &  0.4000 &-313.0   &      0     &    0 &   0\\
  546.0  &  4.667  &  7.000 & -15.00  &  5.000 & 555.0\\
   12.40  & -2.000 &   7.800  &  3.000 & -11.00  & 27.40\\
 -647.0  & -3.667 & -15.33 &  -5.000 &  16.00 & -636.0 \end{bmatrix},\\
    N_2 &= \begin{bmatrix}    
    2.000  &  6.000 &  -5.250 &  -1.000  &  8.000  & -4.250\\
    2.000 &   6.000  &  3.000 &  -1.000  &  8.000  &  4.000\\
   -2.000 &  -6.000  & -3.000 &   1.000  & -8.000  & -4.000\\
   20.00 & -13.00  &  7.000 & -15.00  &  5.000  &  7.000\\
  -14.20 &  43.00  &  0.6005 &   3.000  &-11.00  &  0.8004\\
    3.667 &   6.000  &  4.667 &  -1.000  &  8.000  &  4.000\end{bmatrix},\\
    N_3 &= \begin{bmatrix}    
 -218.0&   -4.400& 0& 0& 0&  0\\
   -4.400& -214.3&  0&  0& 0& 0\\
  -8.250&    1.400&      -8.000&     -7.000&      -16.00&       5.000\\
  212.0&  4.667&     -15.00&    -15.00&        5.000&       7.000\\
   53.80&   -2.000&     7.799&      3.000&      -11.00&    0.8002\\
 1093& -3.667&       8.000&      7.000&       16.00&      -5.000\end{bmatrix},\\
    N_4 &= \begin{bmatrix}    
    -216.0&    4.200&   6.750&    -203.0&      -5.000&    -229.4\\
  0& -214.3&    0.4000&  0&  0& 0\\
 0&    0.4000& -313.0& 0& 0&  0\\
    -579.0&  4.667&    7.000&    -592.0&       5.000&    -570.0\\
     -39.00&  -47.00&  7.801&     -29.00&     -11.00&     -69.00\\
  0& -2.667&  1.667& 0& 0& 0\end{bmatrix}.
  \end{align*}
}%
{\footnotesize
\allowdisplaybreaks
  \begin{align*}
    M_1&= \begin{bmatrix}    
     1&0&0&0&0&0\\
     0&0&0&0&0&0\\
     0&0&0&0&0&0\\
     0&0&0&1&0&0\\
     0&0&0&0&1&0\\
    -1&0&0&0&0&0\end{bmatrix},
    M_2 = \begin{bmatrix}    
    0&1&0&0&0&0\\
    0&1&0&0&0&0\\
    0&-1&0&0&0&0\\
    0&0&0&1&0&0\\
    0&0&0&0&1&0\\
    0&1&0&0&0&0\end{bmatrix},\\
    M_3 &= \begin{bmatrix}    
    0&0&0&0&0&0\\
    0&0&0&0&0&0\\
    0&0&1&0&0&0\\
    0&0&0&1&0&0\\
    0&0&0&0&1&0\\
    0&0&-1&0&0&0\end{bmatrix},
    M_4 = \begin{bmatrix}    
    0&0&0&-1&0&0\\
    0&0&0&0&0&0\\
    0&0&0&0&0&0\\
    0&0&0&1&0&0\\
    0&0&0&0&1&0\\
    0&0&0&0&0&0\end{bmatrix}.
  \end{align*}
}
{\footnotesize  
  \begin{align*}
    L_1 &= \begin{bmatrix}    
     1&17& 2\\
     0& 0& 0\\
     0& 0& 0\\
   -13&-15&7\\
    43&15& 8\\
    -1&-17&-2\end{bmatrix},
    L_2 = \begin{bmatrix}    
      -3&  2&  4\\
      -3&  2&  4\\
       3& -2& -4\\
     -20& -2&  7\\
      14& -7&  8\\
      -3&  2&  4\end{bmatrix},\\
    L_3 &= \begin{bmatrix}    
      3.095\times 10^{-3}&   0& 8.646\times 10^{-4}\\
     6.644\times 10^{-4}& 0& 8.238\times 10^{-4}\\
        -5.001&       1.000&     4.999\\
        -9.002&     -13.00&       7.000\\
         -15.00&      43.00&     7.999\\
         4.987&      -1.000&    -5.004\end{bmatrix},\\
    L_4 &= \begin{bmatrix}    
         2.001&      22.00&      15.00&     -8.999\\
     0& 7.377\times 10^{-4}&  0&  7.435\times 10^{-4}\\
     0& 0& 0& -2.423\times 10^{-4}\\
        -1.996&     -22.00&     -15.00&      9.001\\
        -6.999&      28.00&      15.00&       15.00\\
       0&  0&  0&  0\end{bmatrix}.
  \end{align*}
}%
{\footnotesize 
\allowdisplaybreaks
  \begin{align*}
    H_1 &= \begin{bmatrix}    
        0&    0&    0\\
       1&    0&    0\\
        0&   1&    0\\
        0&    0&    0\\
        0&    0&    0\\
        0&    0&   1 \end{bmatrix},
    H_2 = \begin{bmatrix}
    -1& 1& 0\\
    0& 0& 0\\
   1& 0& 0\\
    0& 0& 0\\
    0& 0& 0\\
    -1& 0& 1 \end{bmatrix},\\
    H_3 &= \begin{bmatrix}    
    1&0&0\\
    0&1&0\\
   0&0&0\\
   0&0&0\\
    0&0&0\\
   -1&0&1\end{bmatrix},
    H_4 = \begin{bmatrix}
    1&1&0&-1\\
   0&1&0&0\\
   0&0&1&0\\
    0&0&0&0\\
   0&0&0&0\\
    0&-1&0&1 \end{bmatrix}.
  \end{align*}
}%
\end{document}